\documentclass[aps,prc,twocolumn,showpacs,preprintnumbers,amsmath,amssymb]{revtex4-1}

\usepackage{amssymb}
\usepackage{graphicx}
\usepackage{bm}
\usepackage{color}
\usepackage{natbib}

\begin{document}

\title{Effects of equation of state on hydrodynamic expansion, spectra, flow harmonics and two-pion interferometry}
\author{Danuce M. Dudek$^{1}$, Wei-Liang Qian$^{2,3}$, Chen Wu$^{4}$, Ot\'avio Socolowski Jr.$^{5}$, Sandra S. Padula$^{6}$, Gast\~ao Krein$^{6}$, Yogiro Hama$^{7}$, Takeshi Kodama$^{8,9}$}

\affiliation{$^1$Universidade Federal Fronteira Sul - Campus Realeza, 85770-000, Realeza, PR, Brazil}
\affiliation{$^2$Escola de Engenharia de Lorena, Universidade de S\~ao Paulo, 12602-810, Lorena, SP, Brazil}
\affiliation{$^3$Faculdade de Engenharia de Guaratinguet\'a, Universidade Estadual Paulista, 12516-410, Guaratinguet\'a, SP, Brazil}
\affiliation{$^4$Shanghai Institute of Applied Physics, 201800, Shanghai, China}
\affiliation{$^5$Instituto de Matem\'atica Estat\'istica e F\'isica, Universidade Federal do Rio Grande, 96201-900 Rio Grande, RS, Brazil}
\affiliation{$^6$Instituto de F\'{\i}sica Te\'orica, Universidade Estadual Paulista, 01140-070 S\~ao Paulo, SP, Brazil}
\affiliation{$^7$Instituto de F\'{\i}sica, Universidade de S\~ao Paulo, 05389-970 S\~ao Paulo, SP, Brazil}
\affiliation{$^8$Instituto de F\'{\i}sica, Universidade Federal do Rio de Janeiro, 21941-972 Rio de Janeiro, RJ, Brazil}
\affiliation{$^9$Instituto de F\'{\i}sica, Universidade Federal Fluminense, 24210-346, Niter\'oi, RJ, Brazil}

\date{May 17, 2018}

\begin{abstract}
We perform an extensive study of the role played by the equation of state in the hydrodynamic evolution of the matter produced in relativistic heavy ion collisions. 
By using the same initial conditions and freeze-out scenario, the effects of different equations of state are compared by calculating their respective hydrodynamical evolution, particle spectra, harmonic flow coefficients $v_2$, $v_3$ and $v_4$ and two-pion interferometry radius parameters.
The equations of state investigated contain distinct features, such as the nature of the phase transition, as well as strangeness and baryon density contents, which are expected to lead to different hydrodynamic responses.
The results of our calculations are compared to the data recorded at two RHIC energies, 130 GeV and 200 GeV.
The three equations of state used in the calculations are found to describe the data reasonably well. 
Differences can be observed among the studied observables, but they are quite small.
In particular, the collective flow parameters are found not to be sensitive to the choice of the equation of state, whose implications are discussed.
\end{abstract}


\maketitle

\section{I. Introduction}
\label{sec:intro}

The equation of state (EoS) of strongly interacting matter plays a major role in the hydrodynamic description of the hot and dense system created in heavy ion collisions~\cite{hydro-review-01,hydro-review-02,hydro-review-04,hydro-review-05}.
It governs how the hydrodynamic evolution transforms the initial state conditions and fluctuations into final state femtoscopic correlations and anisotropies in terms of collective flow.
Traditionally, EoS's with a first-order transition between the quark-gluon plasma (QGP) and hadronic phases were employed in the hydrodynamic models.
However, lattice QCD calculations~\cite{lattice-01,lattice-02,lattice-03,lattice-04,lattice-05,lattice-06} indicated that the quark-hadron transition is a smooth crossover with zero baryon density and large strange quark mass.
Recently, by using the HISQ/tree and asqtad actions, the HotQCD Collaboration found $T_c=154(9)$ MeV~\cite{lattice-17}, while by employing a stout-improved staggered fermionic action, $T_c=151(3)(3) $GeV is obtained by the Wuppertal-Budapest Collaboration~\cite{lattice-14,lattice-15}.
This motivated many different equations of state (EoS's) which fit to the lattice data at high temperature, while combining with appropriate properties at low temperature~\cite{eos-pasi-01,eos-pasi-02,eos-pasi-03,eos-latt-01,eos-latt-02,eos-latt-03,eos-latt-04,eos-latt-07,eos-hrg-02,eos-latt-13}.

The assumption of zero baryon density is a reasonable approximation for the initial conditions (IC) of the systems created at RHIC and LHC.
However, strongly interacting matter possesses several conserved charges, such as electric charge, net baryon number and strangeness.
Studies have shown~\cite{phase-07,phase-08,phase-01,phase-02,phase-03} that the thermodynamic properties and phase transitions are modified when the number of conserved charges in the system changes. 
In the case of a liquid-gas phase transition, for instance, the increase in the number of conserved charges increases the dimension of the binodal surface, and the corresponding transition is continuous~\cite{phase-06,phase-04,phase-05}.
In view of this, one can expect that in the case of the QCD matter, conserved charges could affect the duration of the hydrodynamic evolution of the system in the transition region and likely would manifest themselves at the stage of hadronization. 
Therefore, the experimental data on multiplicity, the ratio of particle yields and their fluctuations need to be analyzed through models which handle correctly finite baryon density and strangeness. 
Statistical models with finite chemical potential are capable of describing the data reasonably well~\cite{statistical-model-01,statistical-model-02,statistical-model-03,statistical-model-04}, indicating that it is important for the study of the system evolution to use EoS's providing a reasonable description of the matter produced over a large range of densities and temperatures.
In particular, a variety of model calculations~\cite{qcd-phase-01,qcd-phase-02,qcd-phase-03,qcd-phase-04,qcd-phase-05} indicated the existence of a first order phase transition at non-vanishing chemical potential.
Following this line of thought, a compromise was proposed by Hama et al.~\cite{sph-eos-1}, where a phenomenological critical point is introduced to smoothen the transition region where the baryon density is smaller than that of the critical point.
In the model, finite baryon chemical potential is taken into consideration in both the QGP and the hadronic phase.
Such phenomenological approach reflects well the main characteristic of a smooth crossover transition while explicitly considering non-zero baryon density. 
Unfortunately, in the QGP phase, the model does not accurately reproduce asymptotic properties of the QGP matter.

The present work employs different EoS's in an ideal hydrodynamical model to study their effects on particle spectra, flow harmonics and two-pion interferometry. 
In the following section, we briefly review different EoS's suggested in the literature and then discuss their effects as employed in the present work. 
In section~\ref{sec: num} we present the results of our hydrodynamical simulations, particle spectra, harmonic coefficients as well as two pion interferometry.
The calculations are done for RHIC energies of 130~GeV and 200~GeV, for various centrality windows. 
Conclusions and perspectives for future work are presented in section~\ref{sec: concl}.

\section{Equation of State and Hydrodynamical model}
\label{sec: eos}

Many different EoS's compatible with results of lattice QCD simulations have been investigated in the literature. 
Huovinen~\cite{eos-pasi-01} proposed an EoS connecting a lattice QCD EoS to another one for a hadronic resonance gas (HRG) model, requiring continuity of the entropy density and its derivatives in the transition region. 
In a later work, Huovinen and Petreczky~\cite{eos-pasi-02} improved the parameterization of Ref.~\cite{eos-pasi-01} by focusing on the trace anomaly, $\Theta \equiv {T^\mu}_\mu = e - 3P $. In this case, the lattice EoS at high
temperature is adopted and is connected smoothly to an EoS of an HRG model at low temperature by requiring that the trace anomaly, as well as its first and second derivatives, are continuous.
In Refs.~\cite{eos-latt-01,eos-latt-08}, an EoS was proposed also based on the lattice data and a HRG model. 
In this EoS, the sound velocity is interpolated in the transition region and is constrained by means of thermodynamical relations to match the lattice QCD entropy density through an integral in temperature.
A few other EoS's were proposed along similar lines~\cite{eos-latt-02,eos-latt-03,eos-latt-07}, using the lattice EoS at high temperatures and connecting it to a phenomenological hadronic EoS using different prescriptions. 
In some of those approaches, however, there are issues of thermodynamic consistency.

On the other hand, instead of interpolating lattice QCD data, some works focused on EoS's with a critical end point in the phase diagram. 
In Ref.~\cite{sph-eos-1}, for instance, a phenomenological critical point is introduced via an EoS from the MIT bag model for the QGP phase, connected to an HRG EoS for the hadronic phase. 
Another attempt was implemented in Ref.~\cite{PNJL-01}, where an SU(3) Polyakov-Nambu-Jona-Lasinio (PNJL) model was used for the high-temperature phase.
A critical end point is naturally obtained by using the Polyakov loop as the order parameter of the deconfinement transition.

\begin{figure}[!htb]
\begin{center}
\includegraphics[width=9cm]{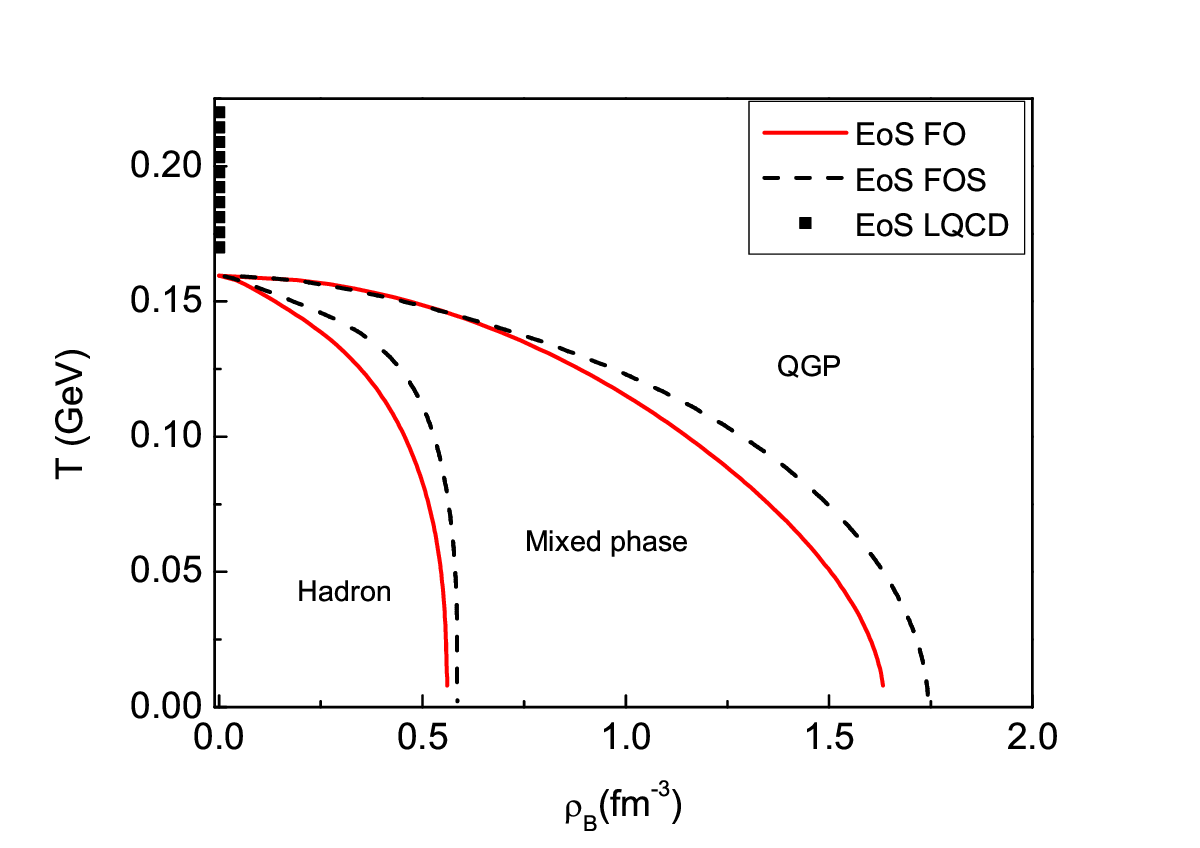} \\
\includegraphics[width=9cm]{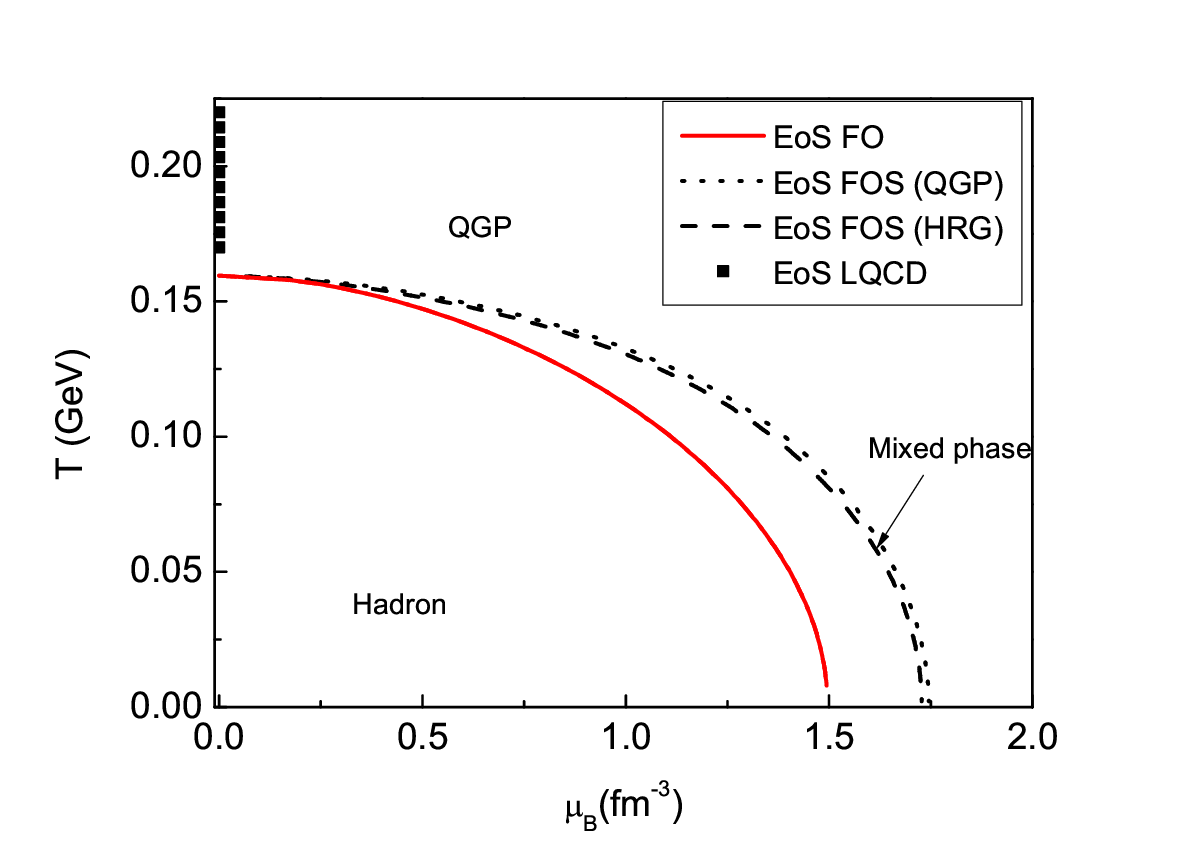}
\end{center}
\caption{(Color online) Phase boundaries of FO and FOS equations of state shown in terms of temperature vs. baryon density (top plot) and in tempetature vs. baryon chemical potential (bottom plot). 
The filled square symbols in the $T$-axis represent the transition region of the LQCD equation of state.}
\label{phase_boundary}
\end{figure}

We note that most of the EoS's discussed above consider only zero baryon density. 
Moreover, in the hydrodynamical simulations, usually averaged IC were used, and only a few works previously adopted full three-dimensional (3-D) hydrodynamical simulations. 
It was estimated in Refs.~\cite{eos-latt-10,eos-pasi-01} that the effect of finite chemical potential is small, less than a few percents.
Owing to baryon density fluctuations, however, it is not clear whether the effect could increase in the case of event-by-event fluctuating IC.
In addition, in most studies, calculations were only done for some specific collision energy and centrality windows. 
In view of this, it seems worthy to carry out an event-by-event 3D simulation on the effects of the EoS, covering a broader range of the published data.

In this work, hydrodynamical calculations are performed by using the full 3-D ideal hydrodynamical code NEXSPheRIO. 
For a more quantitative treatment of heavy ion collisions, the effect of viscosity should be taken into account. 
However, the primary purpose of this study is to qualitatively investigate the differences resulting from using various EoS's rather than to reproduce the data precisely, and viscosity usually reduces such differences.
Besides, viscosity may also introduce additional theoretical uncertainties, such as viscous correction from equilibrium distribution on the freeze-out surface~\cite{hydro-viscosity-fz-01,hydro-viscosity-fz-02}.
The NEXSPheRIO code uses IC provided by the event generator NeXuS~\cite{nexus-1,nexus-rept} and solves the 3+1 ideal hydrodynamic equations with the SPheRIO code~\cite{sph-1st,sph-review-1}.
By generating many NeXuS events, and solving the equations of hydrodynamics independently for each of them, one takes into account the fluctuations of IC on an event-by-event basis.
At the end of the hydrodynamic evolution of each event, a Monte-Carlo generator is employed for hadronization using the Cooper-Frye prescription, the generated hadrons are then fed back to NeXuS where hadronic decay, as well as final state interaction, are handled.
A partial list of referenes describing studies of heavy-ion collisions using the NEXSPheRIO code can be found in Refs.~\cite{sph-v2-1,sph-v2-2,sph-cfo-1,sph-hbt-1,sph-corr-1,sph-corr-ev-4,sph-vn-4}.

In this work, we investigate three different types of~EoS:

\begin{itemize}
\item (LQCD) A lattice QCD EoS proposed by Huovinen and Petreczky~\cite{eos-pasi-02} with zero baryon chemical potential,
\item (CEP) A lattice QCD inspired EoS~\cite{sph-eos-1} with smooth transition and a critical end point, which considers finite baryon chemical potential,
\item (FOS) An EoS with first-order phase transition~\cite{sph-eos-2} which considers both finite baryon chemical potential and local strangeness neutrality. It is a generalization of the EoS (FO) discussed in Ref.~\cite{sph-review-1}.
\end{itemize}

\begin{figure}[!htb]
\centerline{\includegraphics[width=9cm]{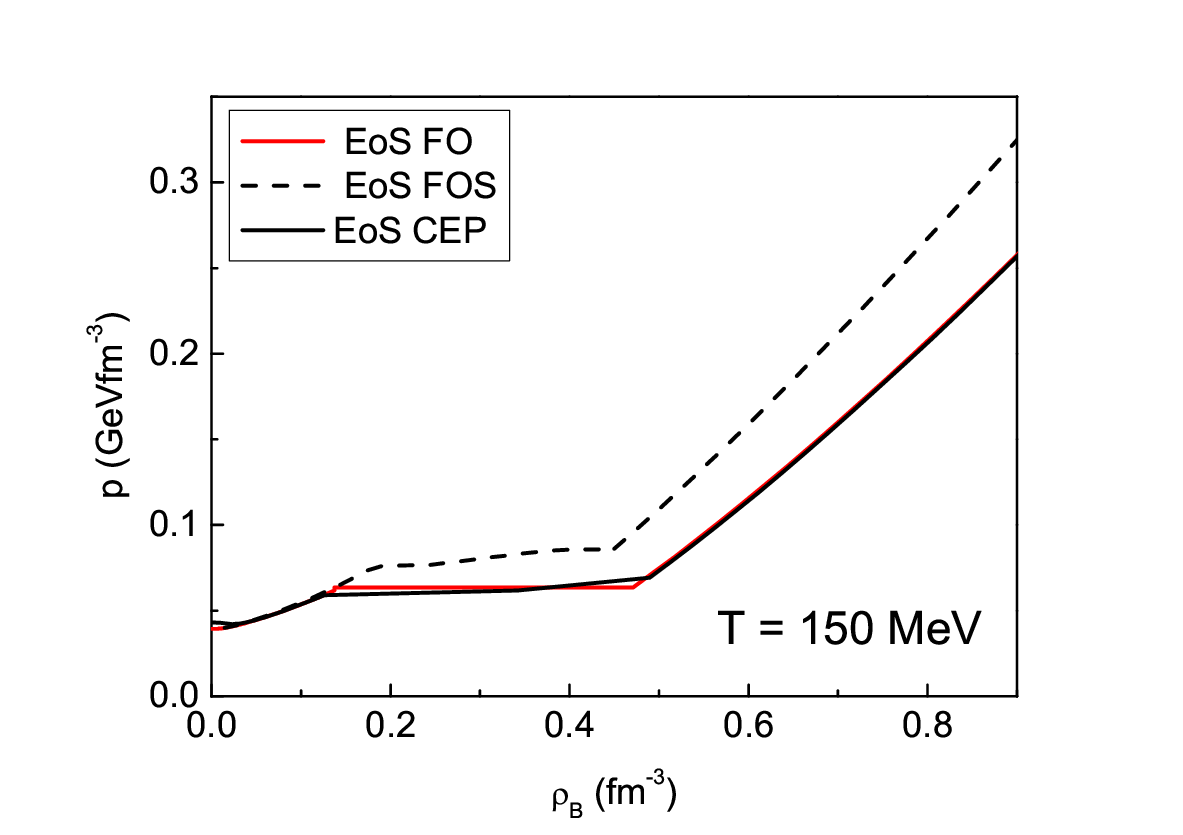} }
\centerline{\includegraphics[width=9cm]{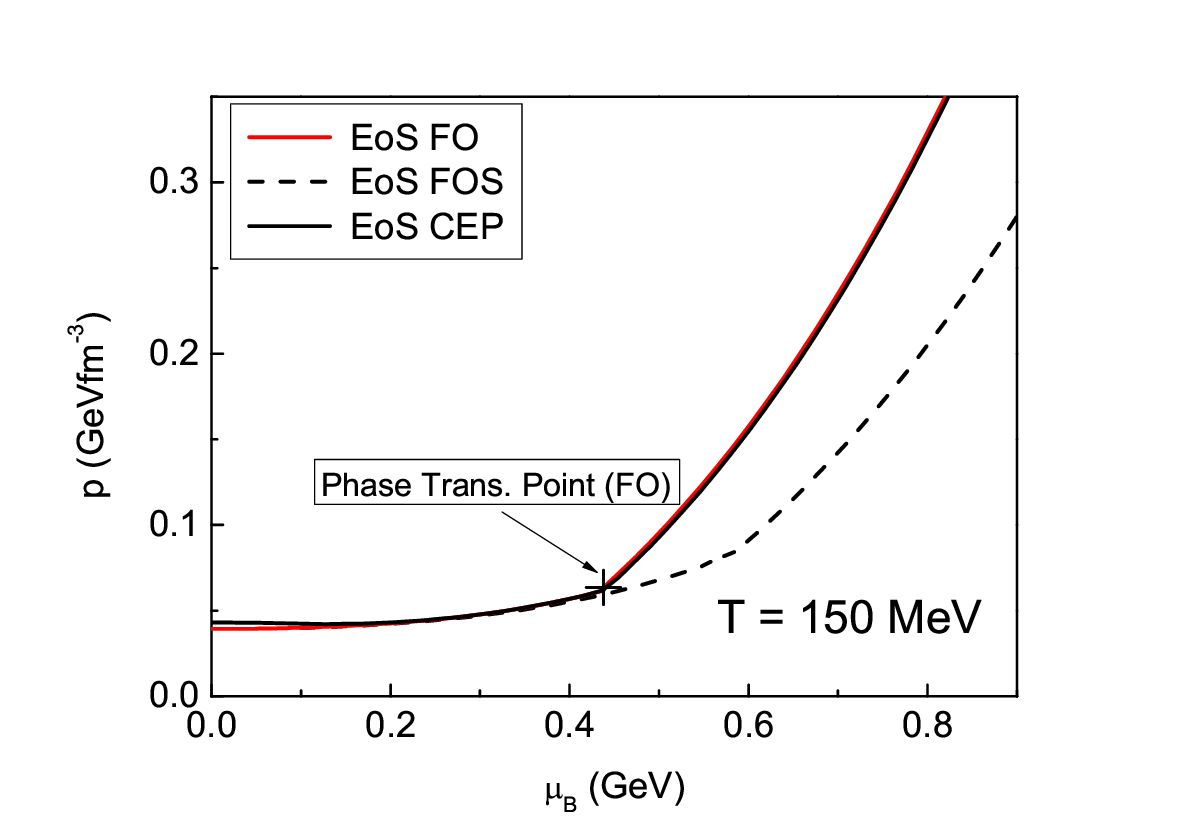} }
\caption{(Color online) Pressure as a function of both the baryon density (top plot) and the baryon chemical potential (bottom plot) at a given temperature $T = 150$ MeV for CEP, FO, and FOS equations of state. The phase transition point in the case of FO is marked on the plot.}
\label{prho}
\end{figure}

\begin{figure*}[htb]
\centerline{
\includegraphics*[width=9cm]{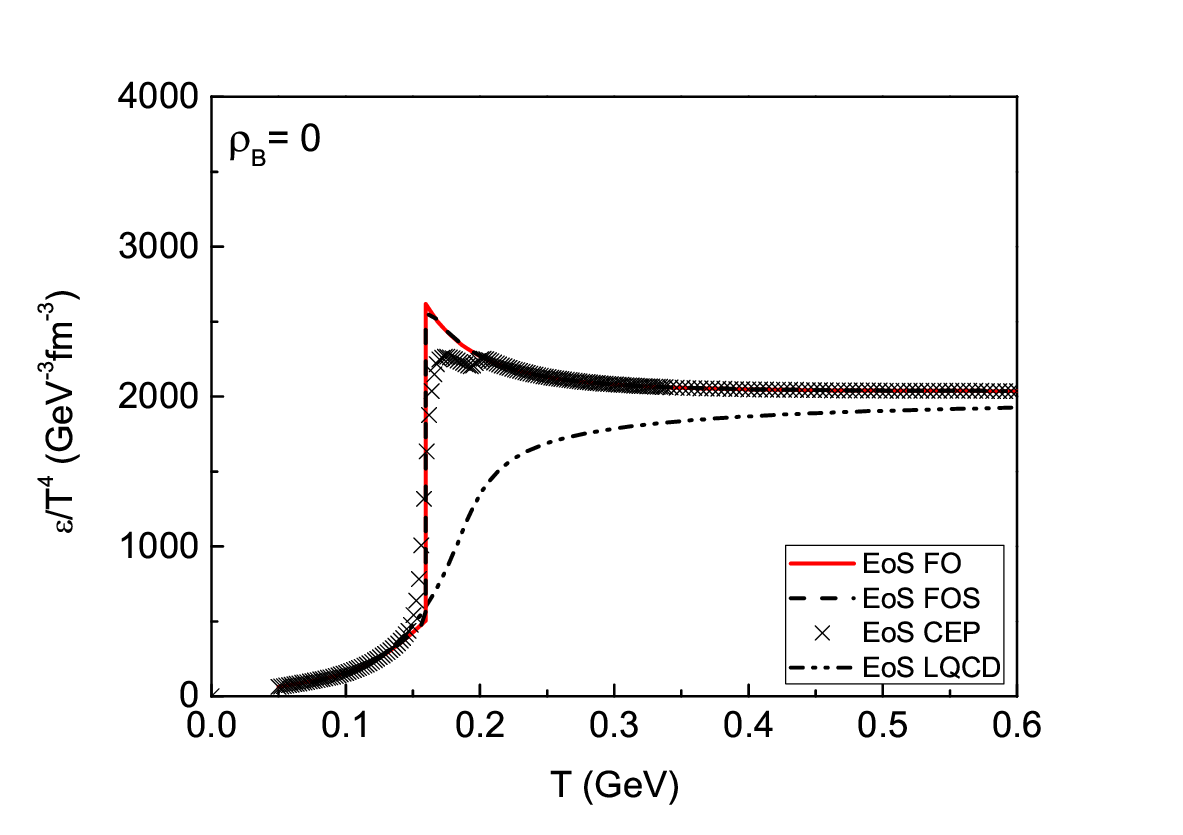}
\includegraphics*[width=9cm]{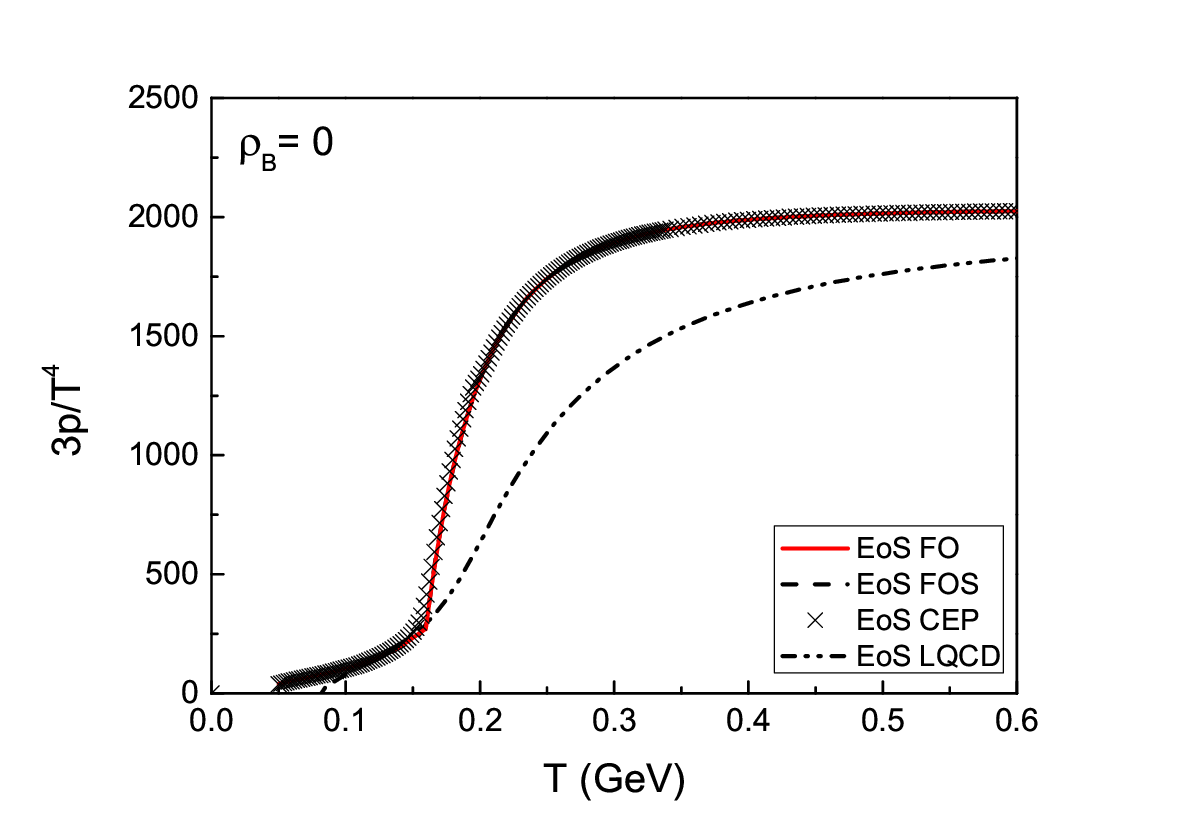} }
\centerline{
\includegraphics*[width=9cm]{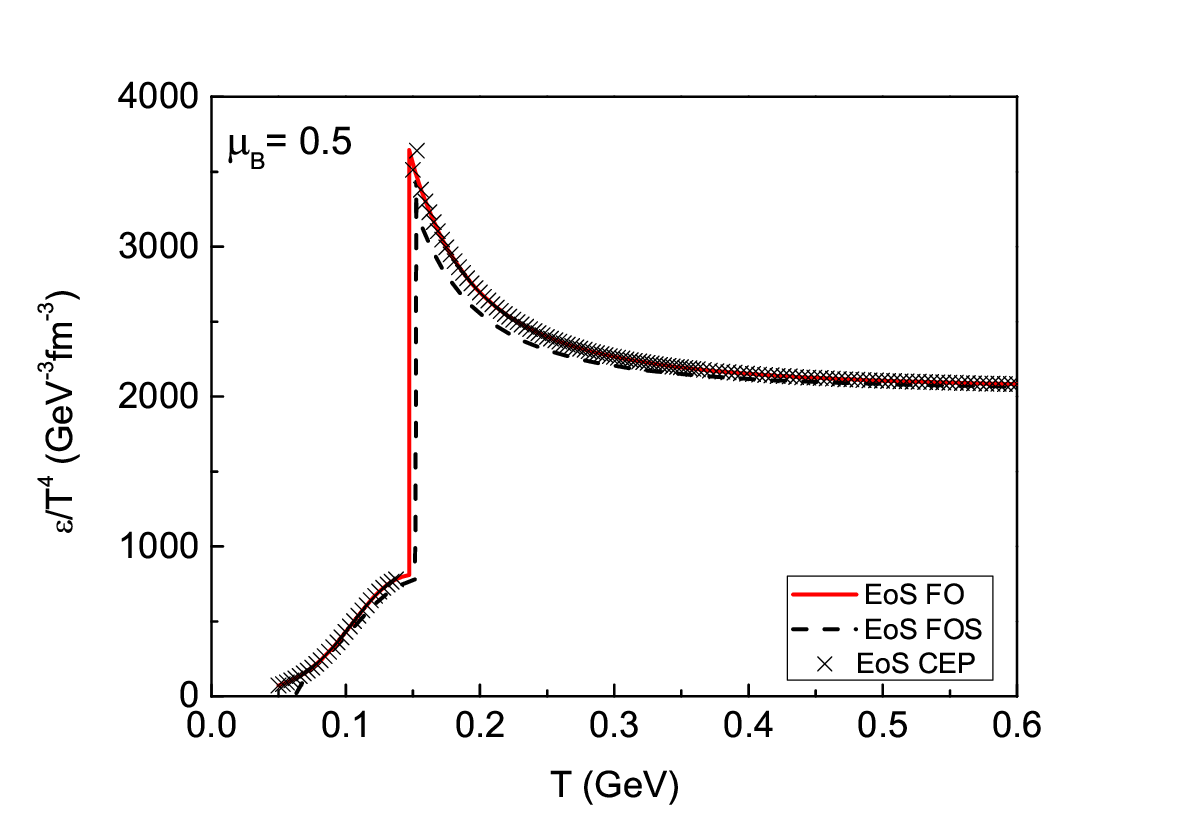}
\includegraphics*[width=9cm]{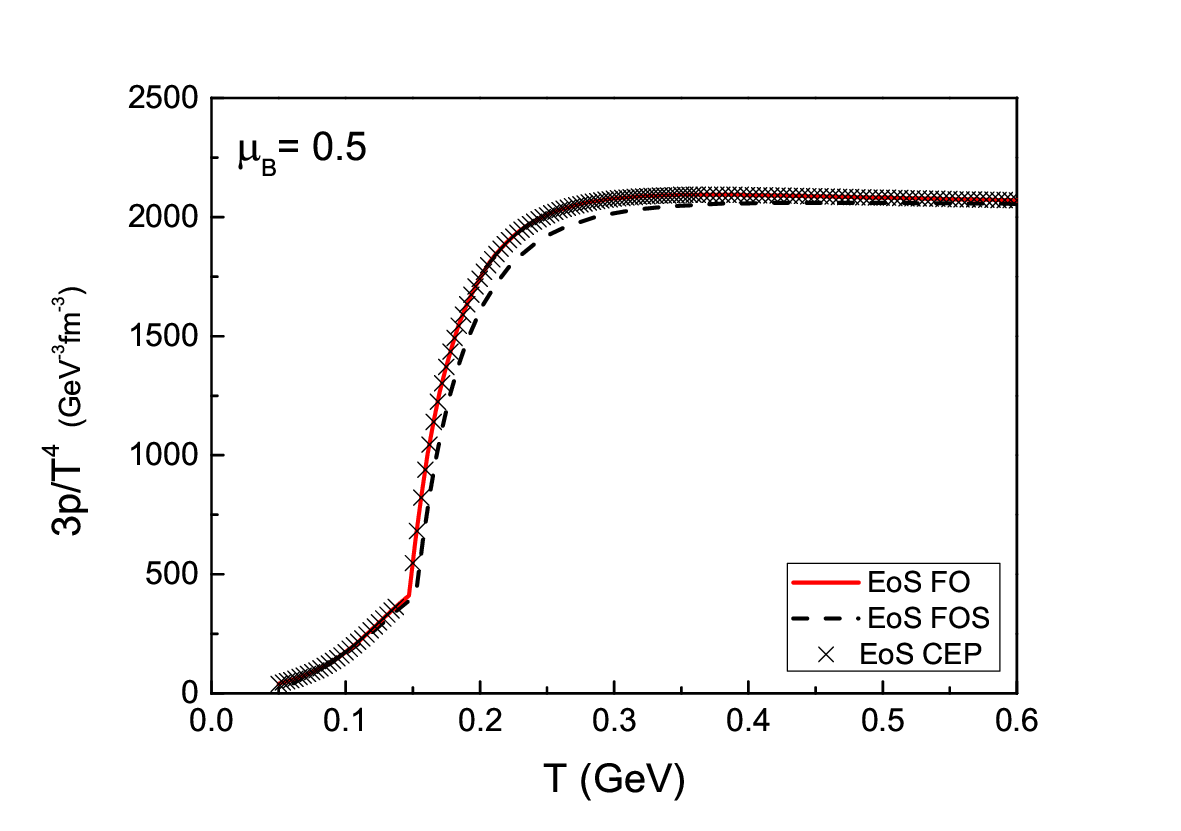} }
\centerline{
\includegraphics*[width=9cm]{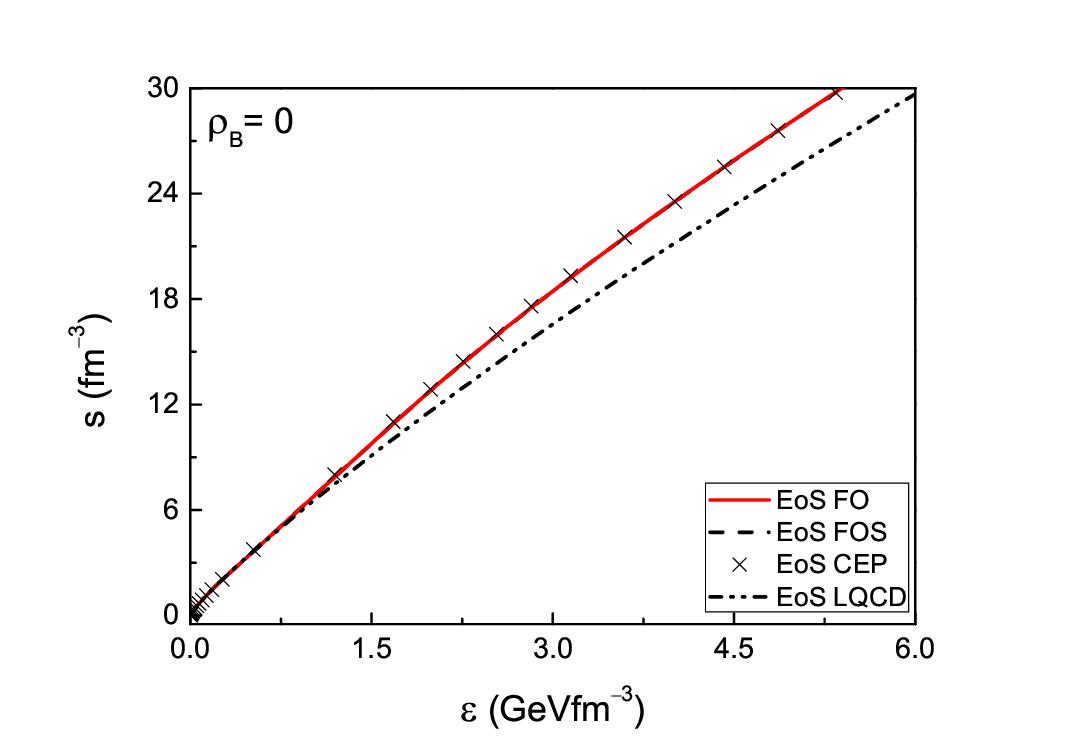}
\includegraphics*[width=9cm]{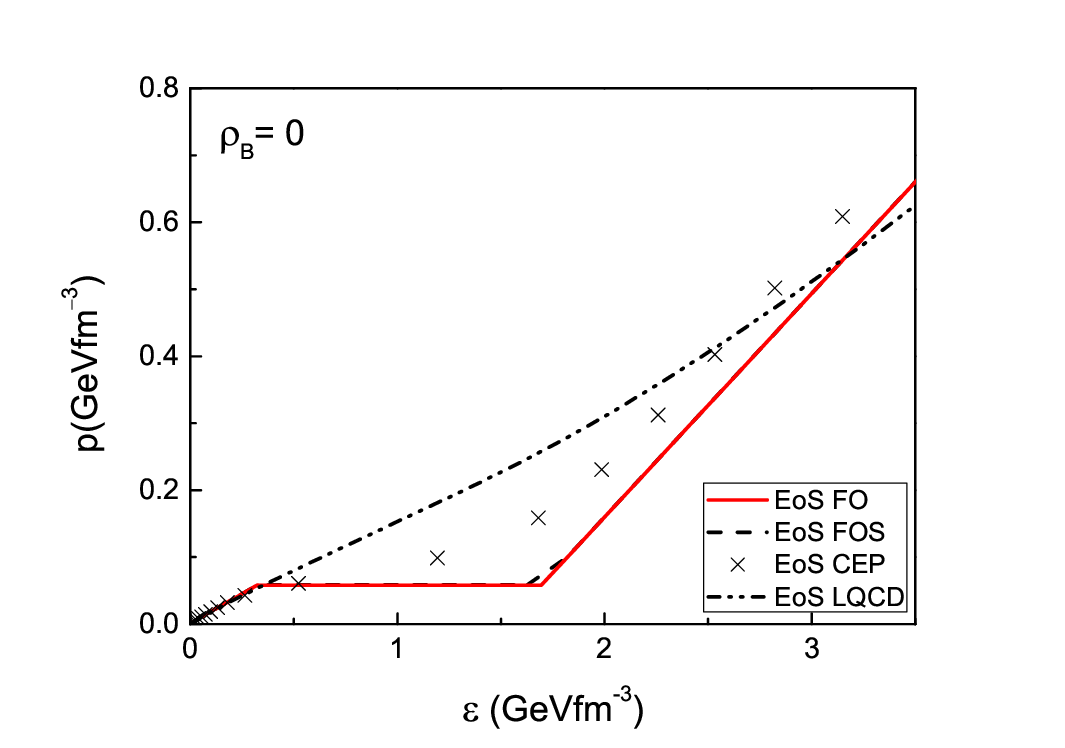} }
\caption{(Color online) Comparison of results using different EoS's: the top panel shows $\epsilon/T^4$ and $3p/T^4$ vs. temperature at zero baryon density; whereas the middle panel shows similar plots at finite chemical baryon potential, for EoS CEP, FO and FOS EoS's; the bottom panel shows entropy density and pressure as a function of the energy density.}
\label{e3pt4}
\end{figure*}
The first EoS, LQCD, adopts the parameterization ``s95p-v1" from~\cite{eos-pasi-02} which fits the lattice QCD data for the high temperature region, while adopting a HRG model for the low temperature region.
This EoS considers vanishing baryon and strangeness densities.
In the calculations, the pressure and energy density are obtained through the trace anomaly $\Theta$ by the following relations
\begin{eqnarray}
\frac{p(T)}{T^4}-\frac{p(T_{low})}{T_{low}^4} &=& \int_{T_{low}}^T \frac{dT'}{{T'}^5}\Theta \, , \nonumber \\
\varepsilon &=& \Theta + 3p \,
\end{eqnarray}
where a sufficiently small lower limit of integration, $T_{low}$, is used in practice~\cite{eos-pasi-02}.
\begin{figure*}[t]
\centerline{
\includegraphics*[width=0.5\textwidth]{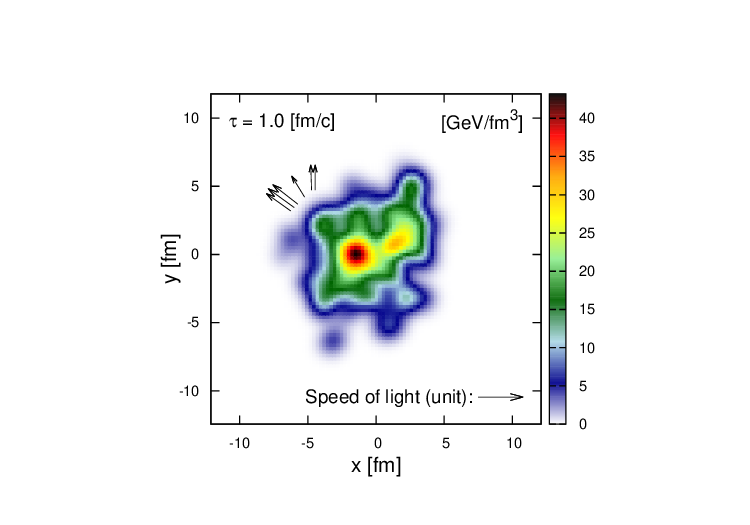}\hspace*{-3cm}
\includegraphics*[width=0.5\textwidth]{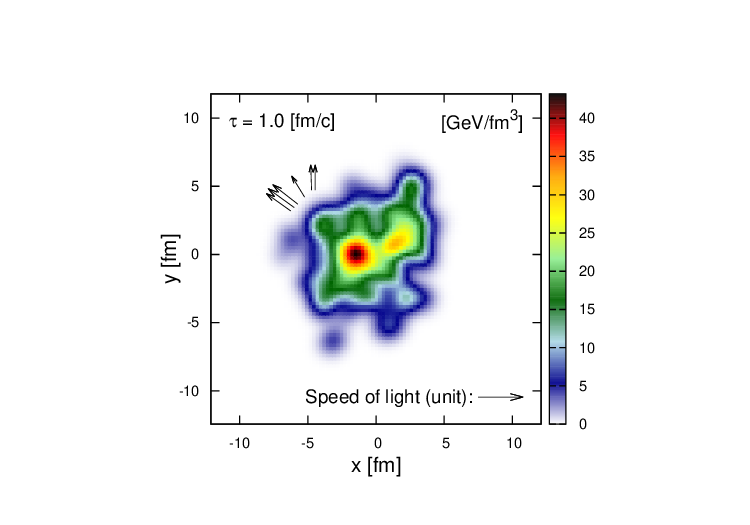}\hspace*{-3cm}
\includegraphics*[width=0.5\textwidth]{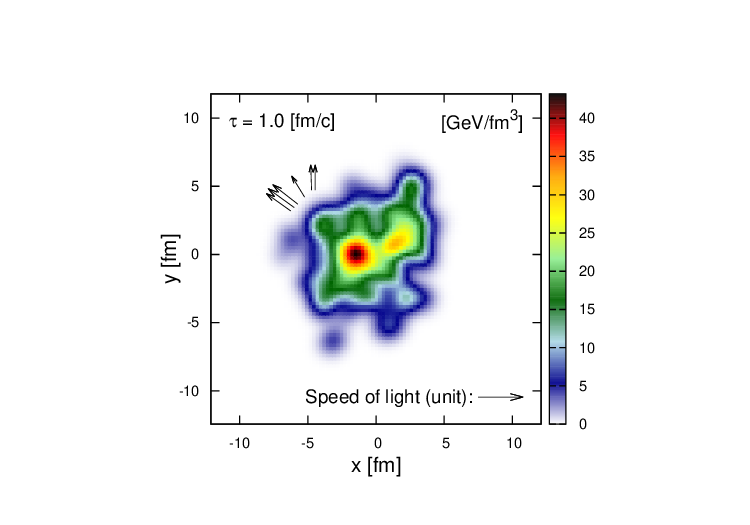}}
\centerline{
\includegraphics*[width=0.5\textwidth]{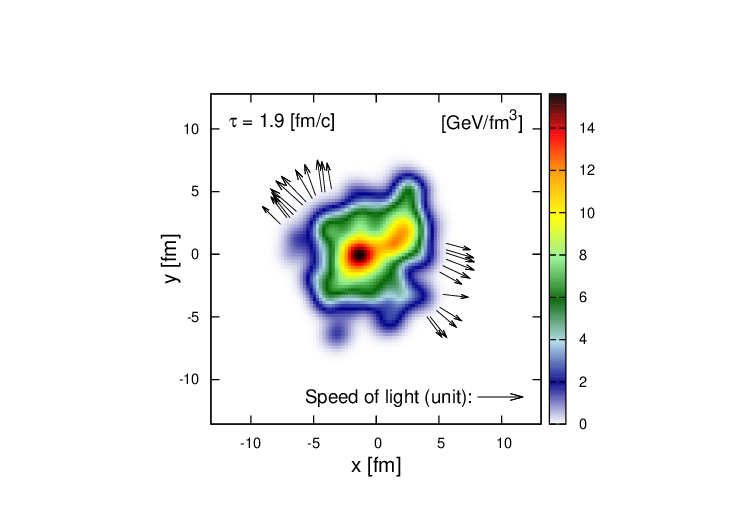}\hspace*{-3cm}
\includegraphics*[width=0.5\textwidth]{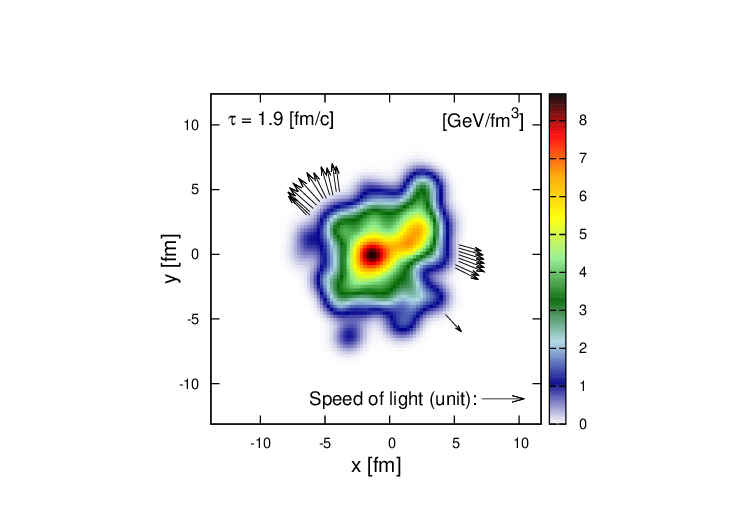}\hspace*{-3cm}
\includegraphics*[width=0.5\textwidth]{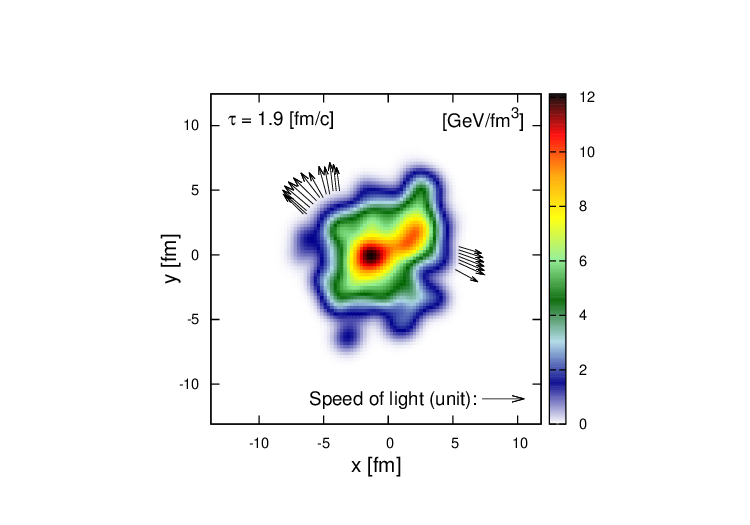}}
\centerline{
\includegraphics*[width=0.5\textwidth]{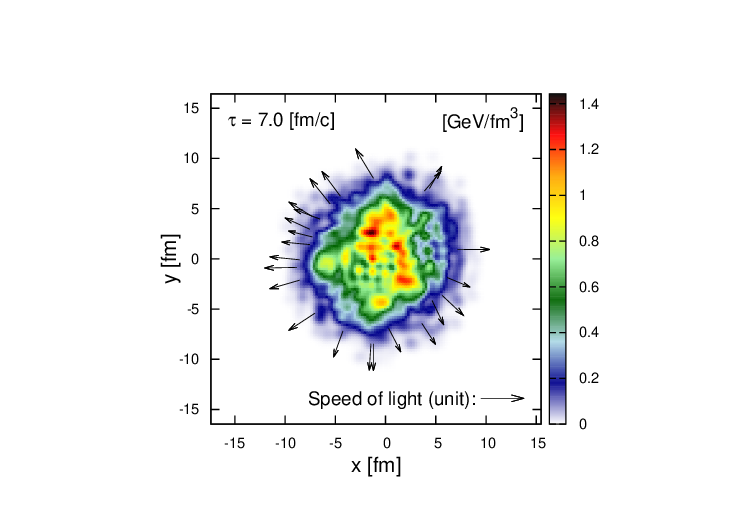}\hspace*{-3cm}
\includegraphics*[width=0.5\textwidth]{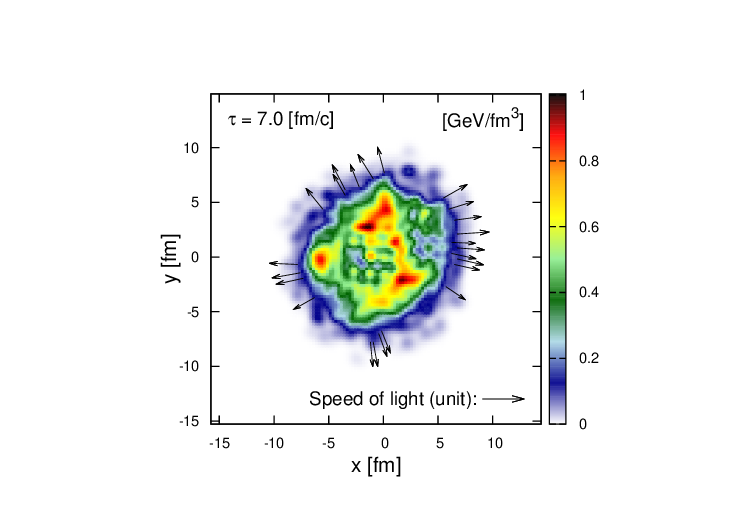}\hspace*{-3cm}
\includegraphics*[width=0.5\textwidth]{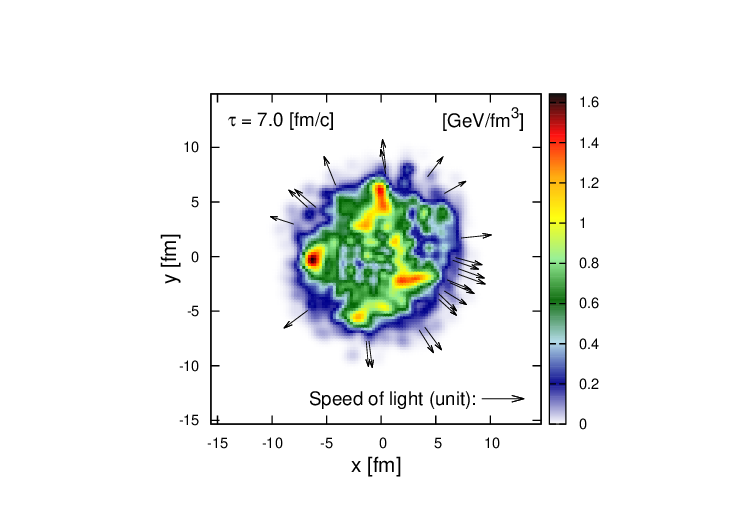}}
\caption{(Color online) Temporal evolution of the energy density profile on the transverse plane $\eta =0$ for a given fluctuating event at 200 GeV using different EoS. 
The arrows show the velocities of the fluid elements on the freeze-out surface. Each panel corresponds to an EoS: LQCD (left column), CEP (middle column) and FOS (right column).}
\label{hydroevolution-eve}
\end{figure*}

The second EoS, CEP, considers the following phenomenological parametrization~\cite{sph-eos-1} instead of Gibbs conditions for the phase transition
\begin{eqnarray}
(p-p^Q)(p-p^H)={\delta}\,, \label{pressure}
\end{eqnarray}
where $p^H$ and $p^Q$ are the pressure in the hadronic and in the QGP phase respectively; $\delta=\delta(\mu_b)$ is a function of baryon chemical potential $\mu_b$ which exponentially approaches zero when $\mu_b$ becomes larger than a critical value $\mu_c = 0.4$ GeV~\cite{sph-eos-1}.
Eq.(\ref{pressure}) has the following solution
\begin{eqnarray}
p  = \lambda p^H+(1-\lambda)p^Q +\frac{2\delta}{\sqrt{(p^Q-p^H)^2+4\delta}}\ , \label{pressure2}
\end{eqnarray}
where
\begin{eqnarray}
\quad \lambda&\equiv&{\frac{1}{2}}\left[1-(p^Q-p^H)/\sqrt{(p^Q-p^H)^2+4\delta}\right]\ .
\end{eqnarray}
When $\mu_b < \mu_c$, Eq.(\ref{pressure2}) gives a smooth transition from the hadronic phase to QGP phase.
On the other hand, it is straightforward to verify that, for small value of $\delta$, $p \rightarrow p^H$ when
$p^Q < p^H$ and $p \rightarrow p^Q$ when $p^Q > p^H$.
As a result, in the region when $\mu_b >\mu_c$, due to the smallness of $\delta$, the phase transition rapidly converges to that of FO~\cite{sph-eos-1}, namely, $p \rightarrow p^Q$ for the QGP phase, $p \rightarrow p^H$ for the hadronic phase, and $p \rightarrow p^Q=p^H$ in the transition region where the Gibbs conditions are satisfied.

The third EoS, FOS, introduces an additional constraint in the FO, namely,
strangeness neutrality, i.e.
\begin{eqnarray}
\rho_s  = 0. \label{rhos0}
\end{eqnarray}
The strangeness chemical potential, $\mu_s$, is introduced in the EoS, not as an independent variable in the system, but for increasing the dimension of the binodal surface of the phase transition~\cite{phase-01}.
It also modifies the phase structure, as discussed below. 
Before carrying out the hydrodynamical simulations, we first examine the differences among the different EoS's qualitatively.

Fig.~\ref{phase_boundary} shows the phase boundaries of the different EoS's.
For LQCD, the deconfinement transition corresponds to the parameterization in the region of $170~{\rm MeV} < T < 220~{\rm MeV}$ on the temperature axis in the plot. 
For FO and FOS, the phase boundary is determined by the Gibbs conditions between the quark-gluon and hadronic phases. 
CEP does not have well-defined phase separation, and therefore it is not shown explicitly in the plot.
Due to the interpolation scheme, it is almost the same as that of FO beyond the critical point (i.e. for $\mu > \mu_c$), and is smoothed out below that point (i.e. for $\mu < \mu_c$).
The top plot shows the phase boundaries in terms of the temperature as a function of baryon density, while the plot in the bottom shows them in terms of the temperature as a function of baryon chemical potential.

We note that FOS possesses a unique feature: the QGP and the hadronic phase boundaries have different baryon chemical potentials. 
It can be seen as a result of the strangeness local neutrality condition. 
This implies that during the phase transition, when the two phases are in equilibrium, it is not necessary that the strangeness density vanishes simultaneously in both phases. 
This is because in the transition region, the strangeness neutrality condition Eq.(\ref{rhos0}) reads,
\begin{eqnarray}
\rho_s  = \lambda \rho_s^H + (1-\lambda) \rho_s^Q = 0.
\label{rhos0tr}
\end{eqnarray}
In other words, in the case of FOS, neither the strangeness density of hadronic phase ($\rho_s^H$) nor that of the QGP phase ($\rho_s^Q$) is necessarily zero in the mixed phase.
Therefore, the baryon chemical potential is not fixed during the phase transition, its value depending on the fraction $\lambda$ of the system in the hadronic phase.
In general, the baryon chemical potential attains different values at the hadronic phase boundary (with $\rho_s^H =0$ and $\lambda = 1$) and at the QGP phase boundary (with $\rho_s^Q=0$ and $\lambda = 0$) as a result of the Gibbs conditions.
In the case of FO, the QGP phase boundary coincides with that of the hadronic phase.

In Fig.~\ref{prho}, the pressure is shown as a function of the baryon density, as well as a function of the baryon chemical potential, for different EoS's at a given temperature $T = 150$ MeV.
It is worth noting here that for FOS, neither the baryon chemical potential nor the strangeness chemical potential is fixed during the isothermal phase transition procedure.
As a result, when expressed in terms of pressure versus the chemical potential, the transition region of FO is a point, but it is a curve in the case of FOS, as shown in the top plot of Fig.~\ref{prho}.
When expressed in terms of pressure versus baryon density, the transition region corresponds to the horizontal line of constant pressure in the case of FO.
On the other hand, the pressure increases during the phase transition in the case of FOS.
Therefore, the phase transition in FOS is smoother than that in FO.
For the CEP EoS, due to its parameterizations, the transition region is smoothed out as compared to that of FO, the pressure also monotonically increases during the process.

\begin{figure}[t]
\begin{center}
\includegraphics[width=9cm]{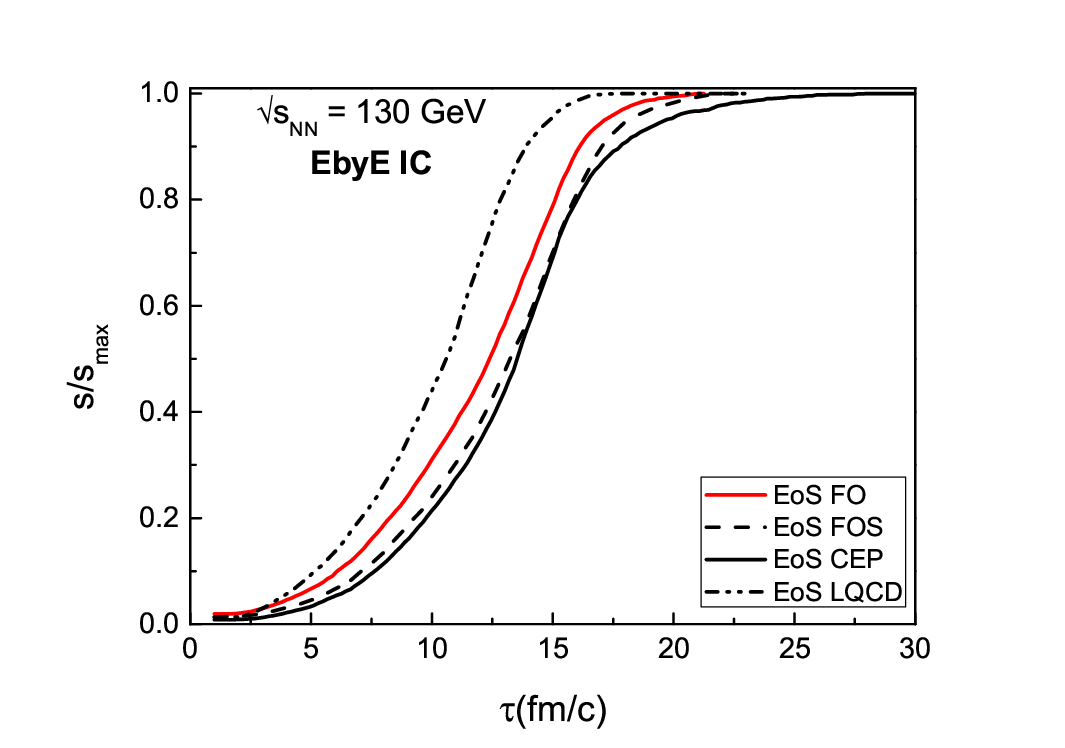} \\
\includegraphics[width=9cm]{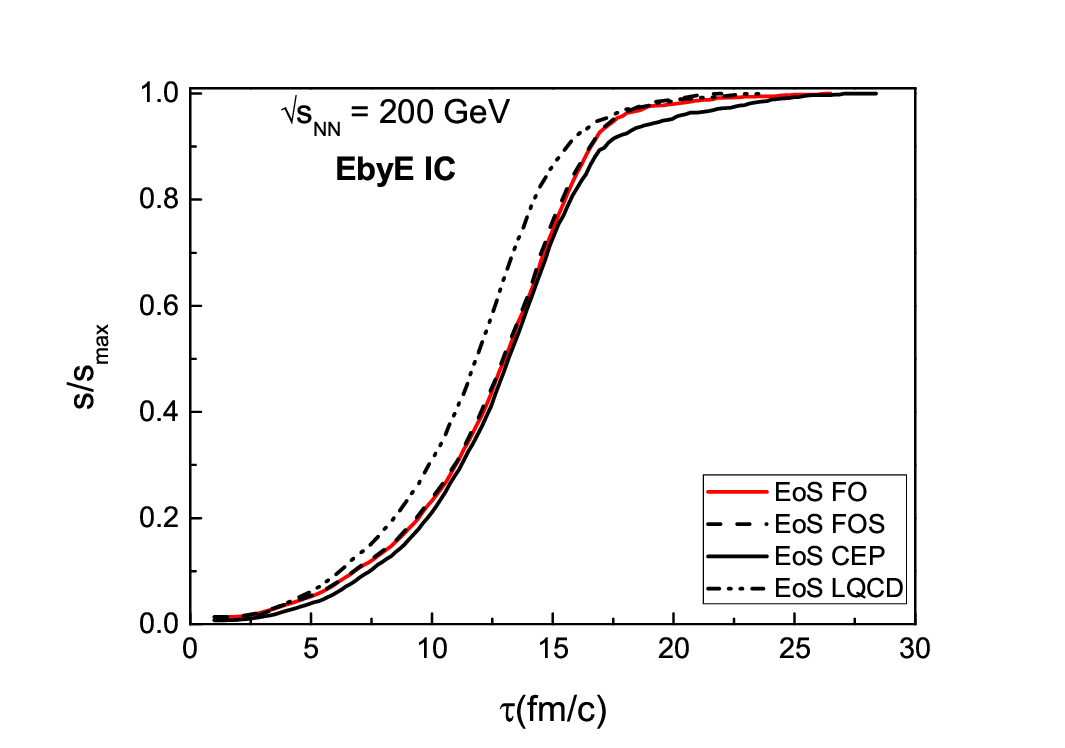}
\end{center}
\caption{(Color online) The freeze-out entropy as a function of the evolution time corresponding to different EoS, is illustrated for a radomly chosen fluctuating event at 130 GeV RHIC energy (top) and 200 GeV (bottom). The entropy is presented in percentage of its initial value.}
\label{entropyevolution}
\end{figure}
The values of $\epsilon/T^4$ and $3p/T^4$ are plotted as a function of the temperature $T$ for all the EoS's in Fig.~\ref{e3pt4}.
In the temperature range 500-600 MeV, the curves for $\epsilon/T^4$ and $3p/T^4$ corresponding to the LQCD EoS have a distinct behavior than the other EoS's, reflecting the fact that it is a fit to the QCD results.
In this region, all the other EoS converge to the non-interacting ideal gas limit.
On the other hand, in the low-temperature limit, all the EoS approach the HRG model, as expected.
The differences between CEP, FO and FOS come from the transition region around the temperature of the first order phase transition $T \sim 160$ MeV employed in FO and FOS.
Since a first order phase transition of a one-component system occurs at a constant temperature, it gives a vertical line in the case of FO.
CEP is smoother in comparison with FO due to its phenomenological parameterization.
Although the strangeness chemical potential is considered in FOS, it gives the same
result as FO.
This can be understood by studying the conditions at the phase boundaries.
At vanishing baryonic chemical potential $\mu_b =0$, it is easy to see that the hadronic phase boundary $\rho_s^H =0$ and the QGP phase boundary $\rho_s^H$ share the same solution, namely, $\mu_s =0$, which corresponds to zero baryon density and strangeness neutrality for both phases.
This is manifested as the intersection between the phase boundaries and y-axis in the bottom plot of Fig.\ref{phase_boundary}.
The middle panel in Fig.\ref{e3pt4} shows the results of CEP, FO and FOS for finite chemical potential (hence finite baryon density). 
The curves at zero chemical potential are also plotted for comparison purposes. 
It can be seen that $\mu_b = 0.5$ GeV, which is beyond the critical point in the case of CEP ($\mu_c=0.4$ is the value assumed here for the chemical potential at the critical point), corresponds to a region resembling the first order phase transition.
Therefore CEP behaves similarly to FO in this case.
On the other hand, FOS is slightly different from them since the corresponding transition is not isothermal at finite baryon density. 
Nevertheless, all three EoS show very similar features in high and low-temperature limits.

\section{Numerical results and discussions}
\label{sec: num}
\begin{figure*}[!htb]
\centerline{
\includegraphics*[width=9cm]{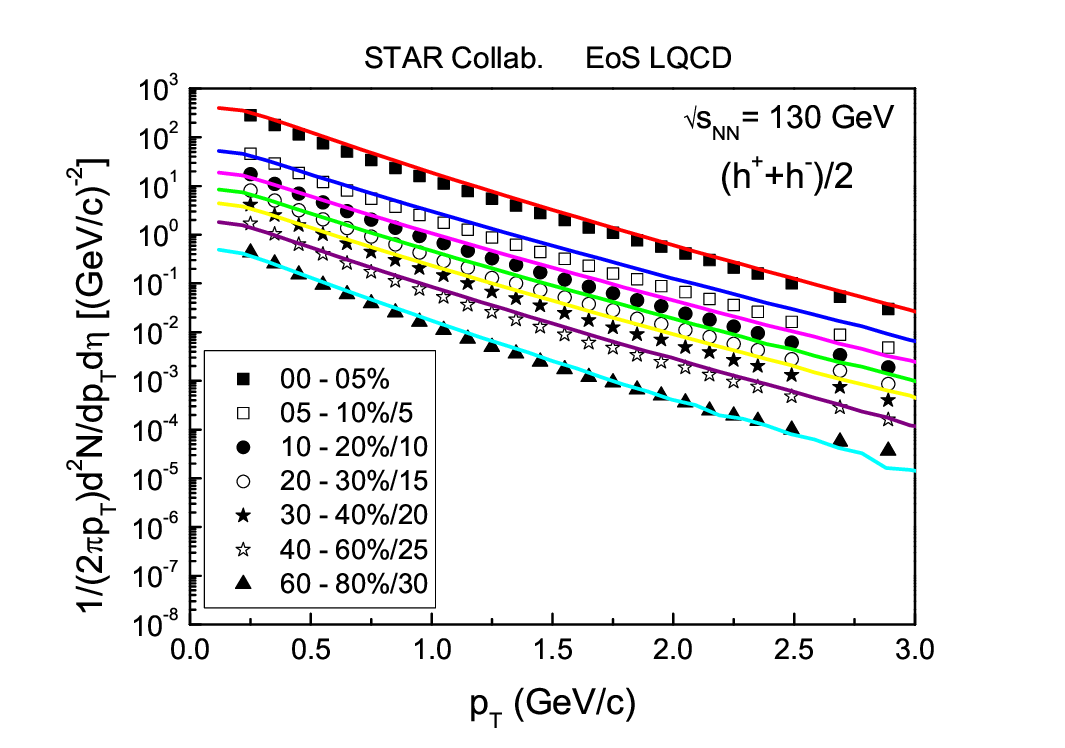}
\includegraphics*[width=9cm]{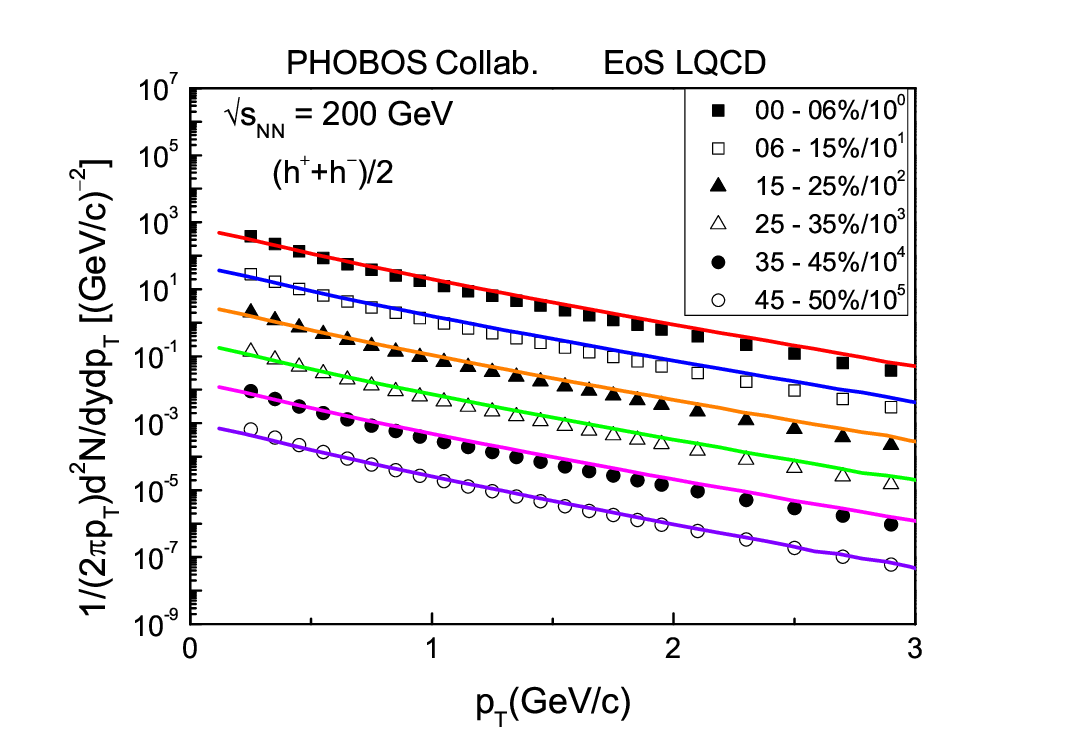} }
\centerline{
\includegraphics*[width=9cm]{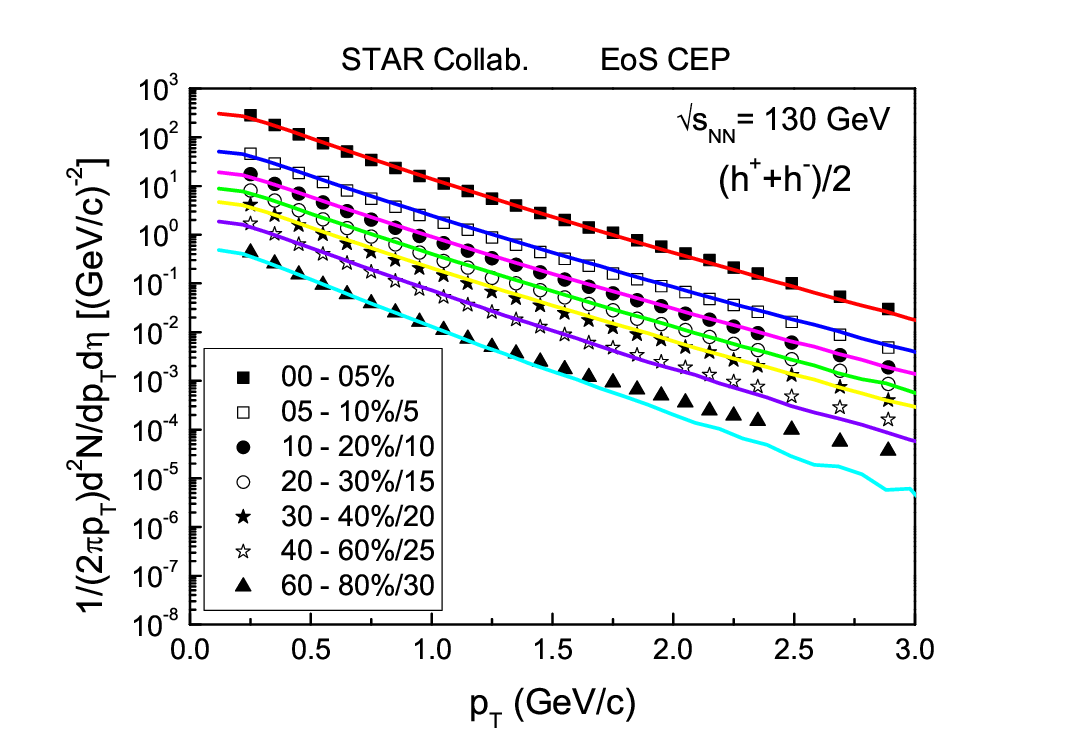}
\includegraphics*[width=9cm]{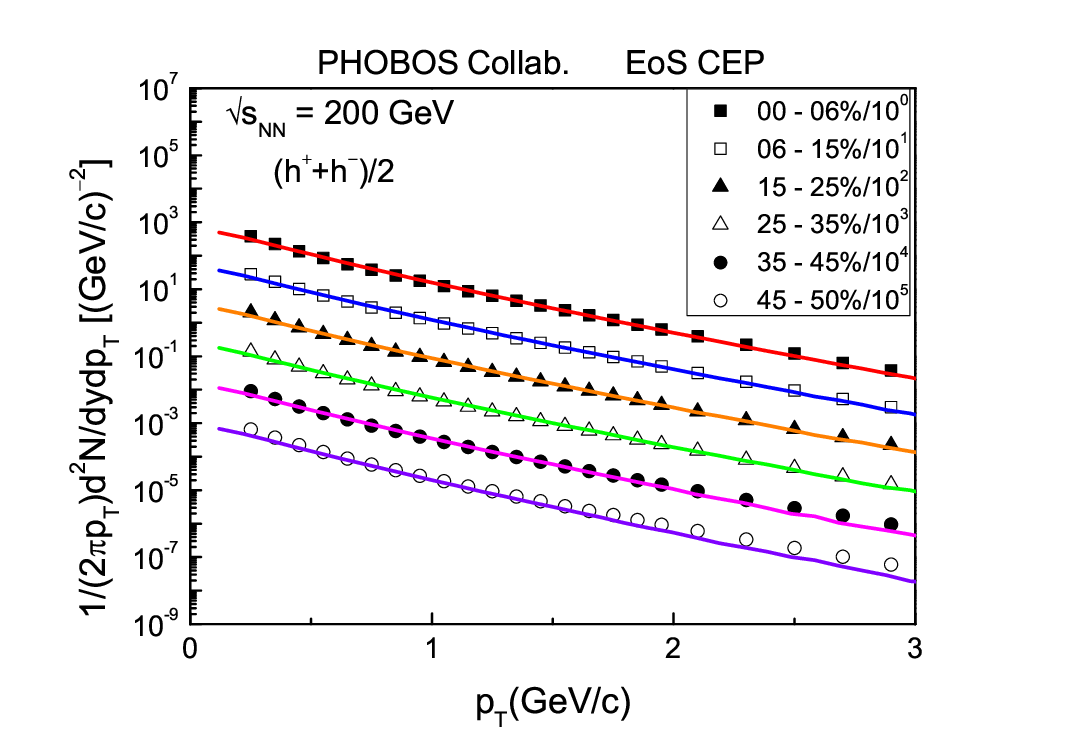} }
\centerline{
\includegraphics*[width=9cm]{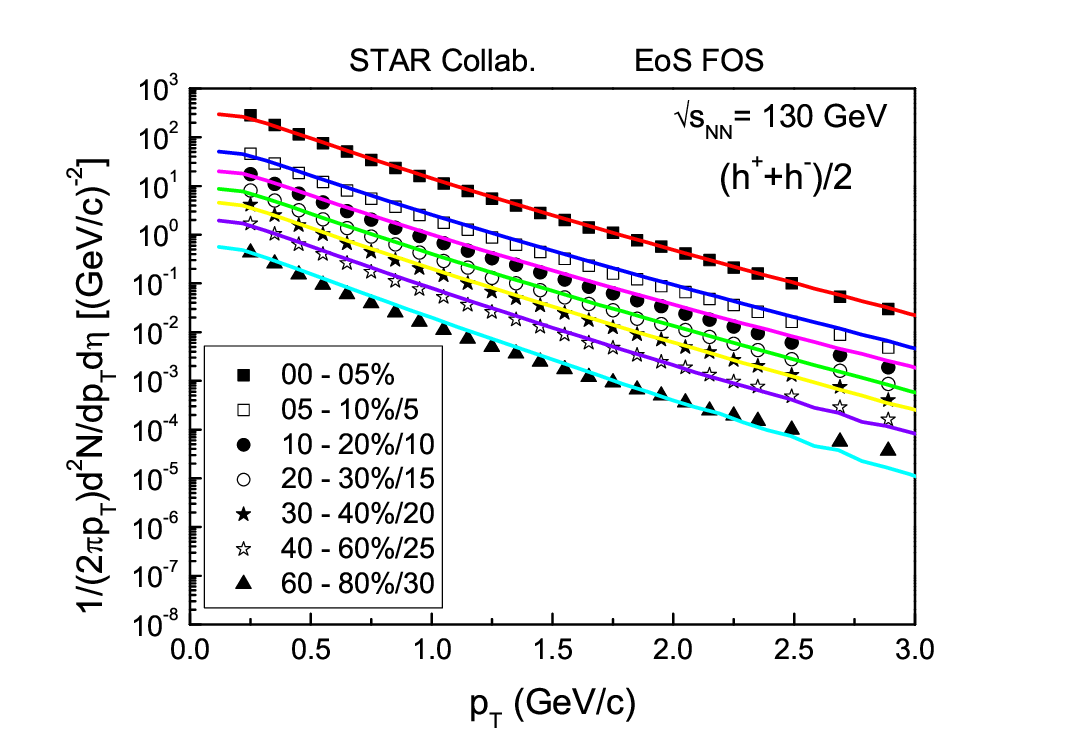}
\includegraphics*[width=9cm]{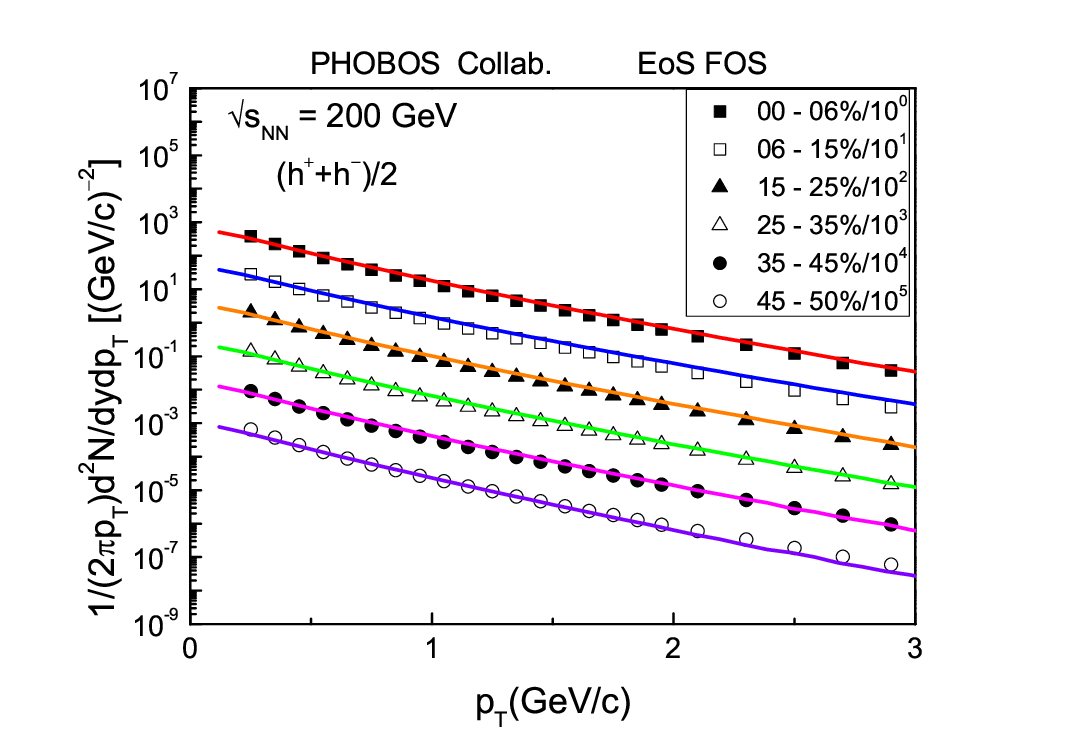} }
\caption{(Color online) The $p_T$ spectra of all charged particles for Au+Au collisions at 130 GeV (left column) and 200 GeV (right column) corresponding to the FOS, CEP and LQCD EoS's.
The data on the left panel is from the STAR Collaboration, whereas in the right panel, are from the PHOBOS Collaboration.}
\label{pt-spectra}
\end{figure*}

\begin{figure*}[!htb]
\centerline{
\includegraphics[width=9cm]{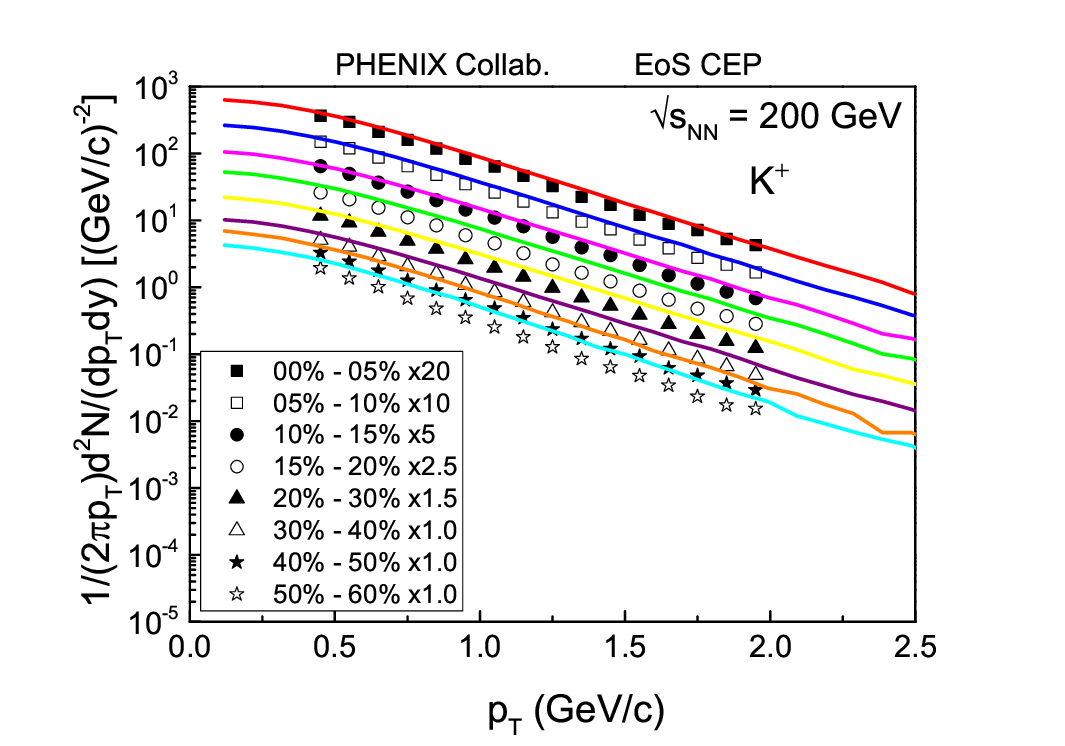}
\includegraphics[width=9cm]{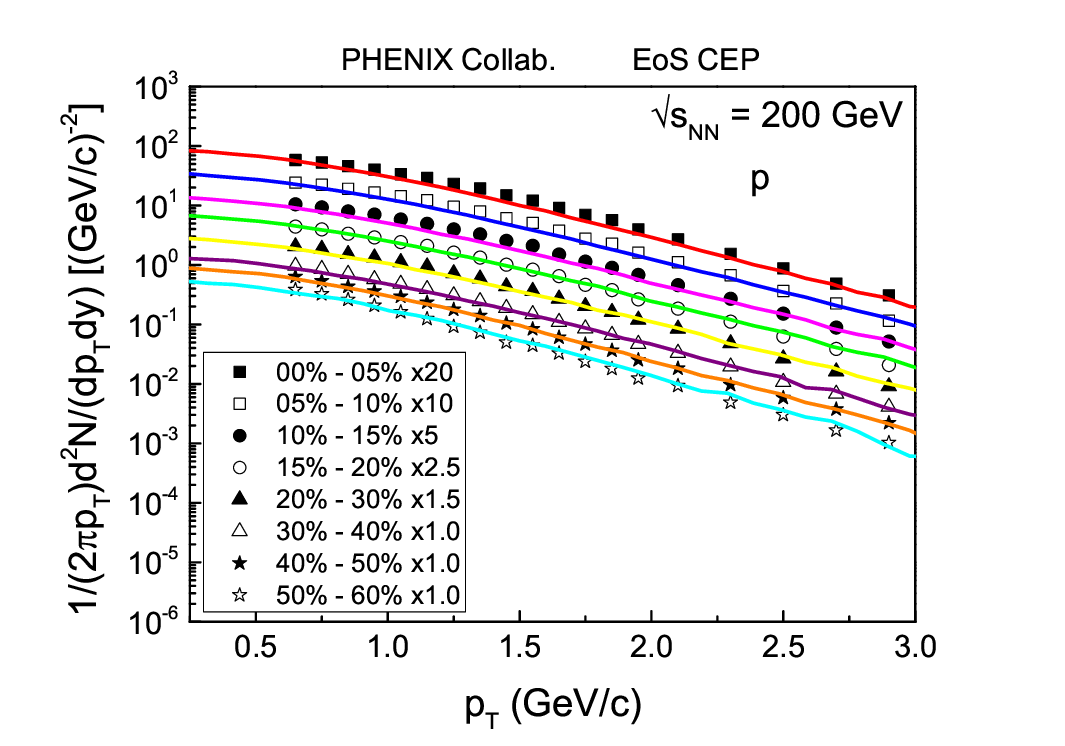}}
\centerline{
\includegraphics[width=9cm]{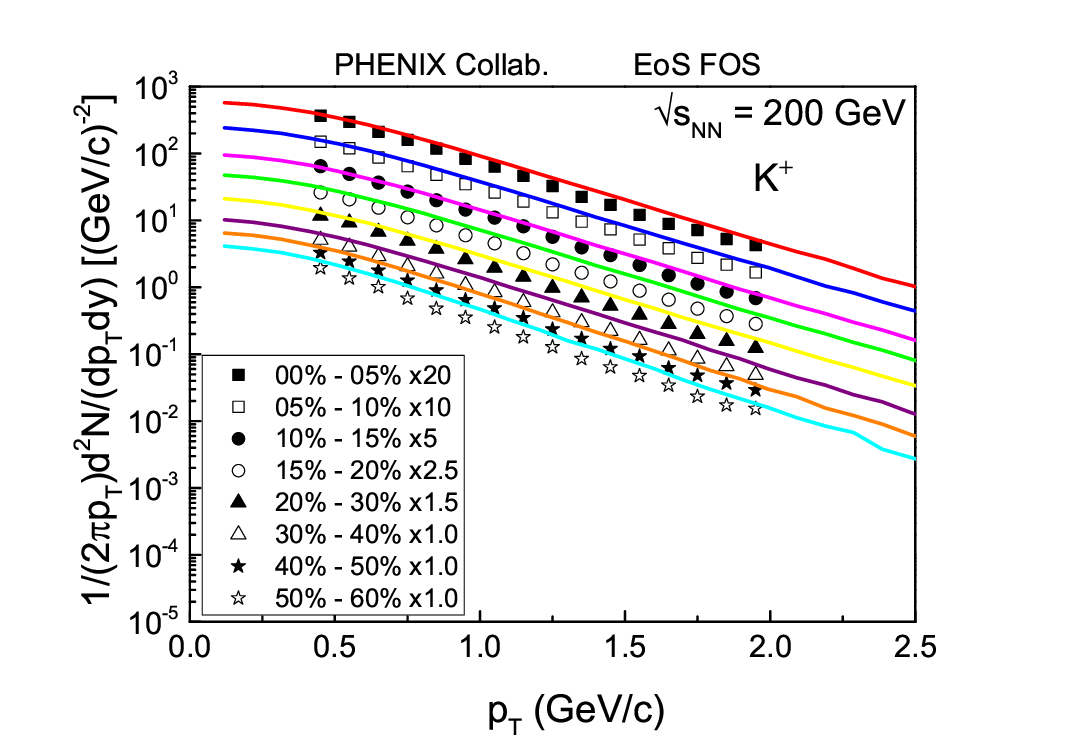}
\includegraphics[width=9cm]{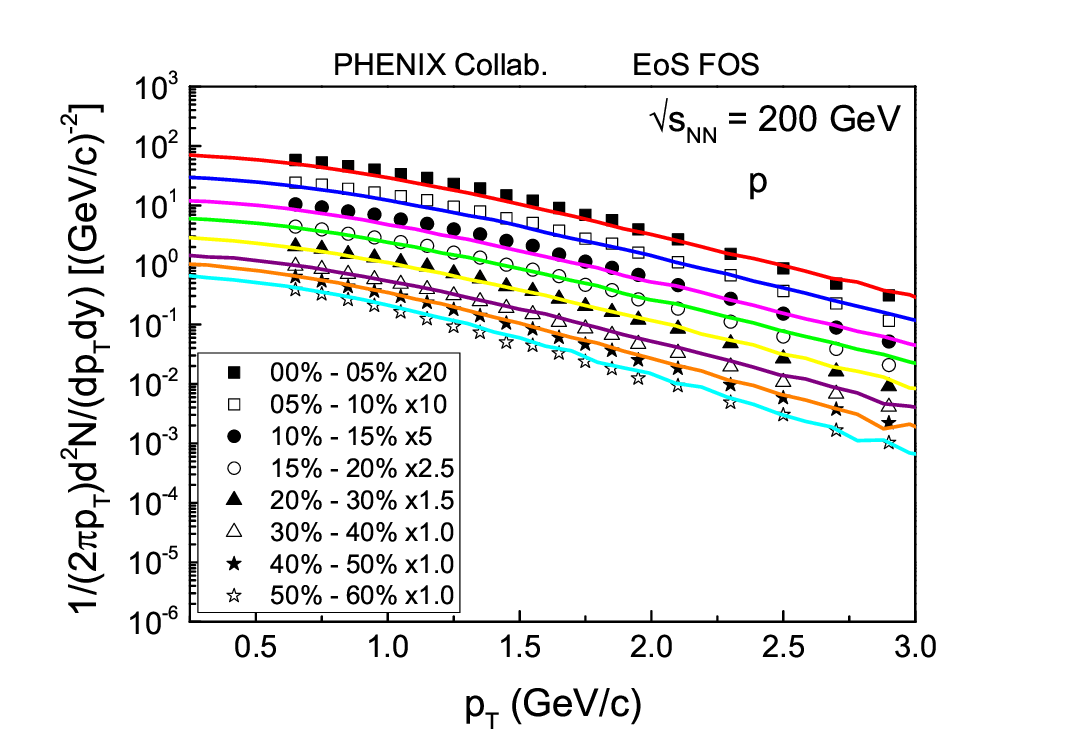}}
\centerline{
\includegraphics[width=9cm]{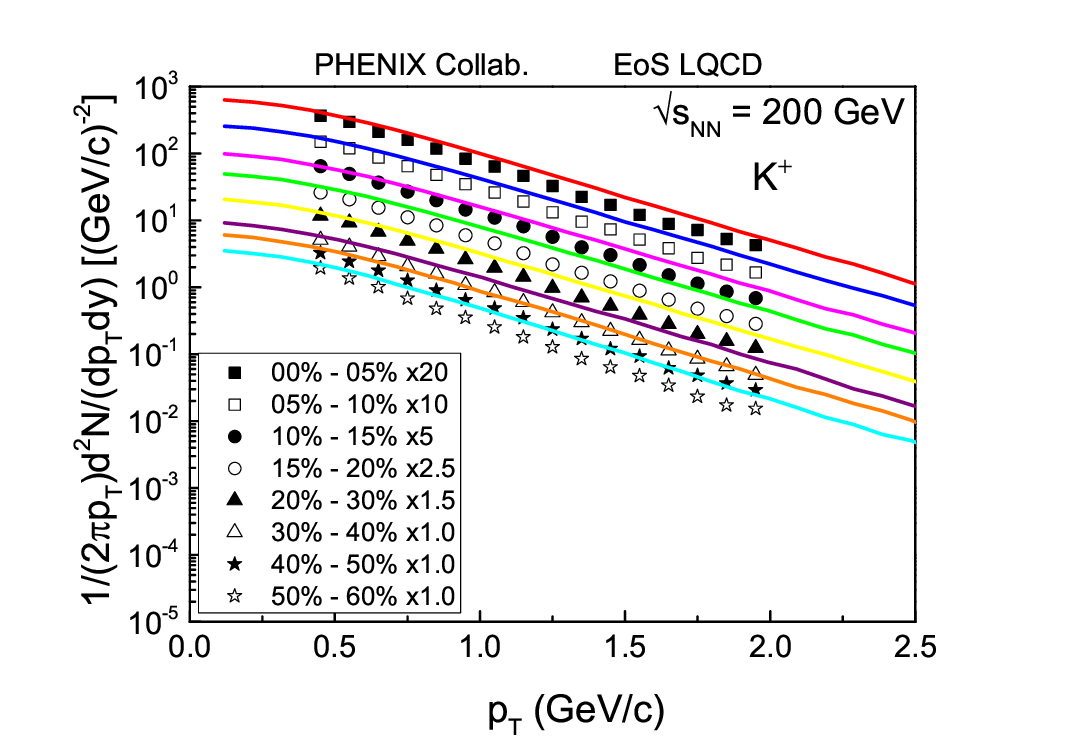}
\includegraphics[width=9cm]{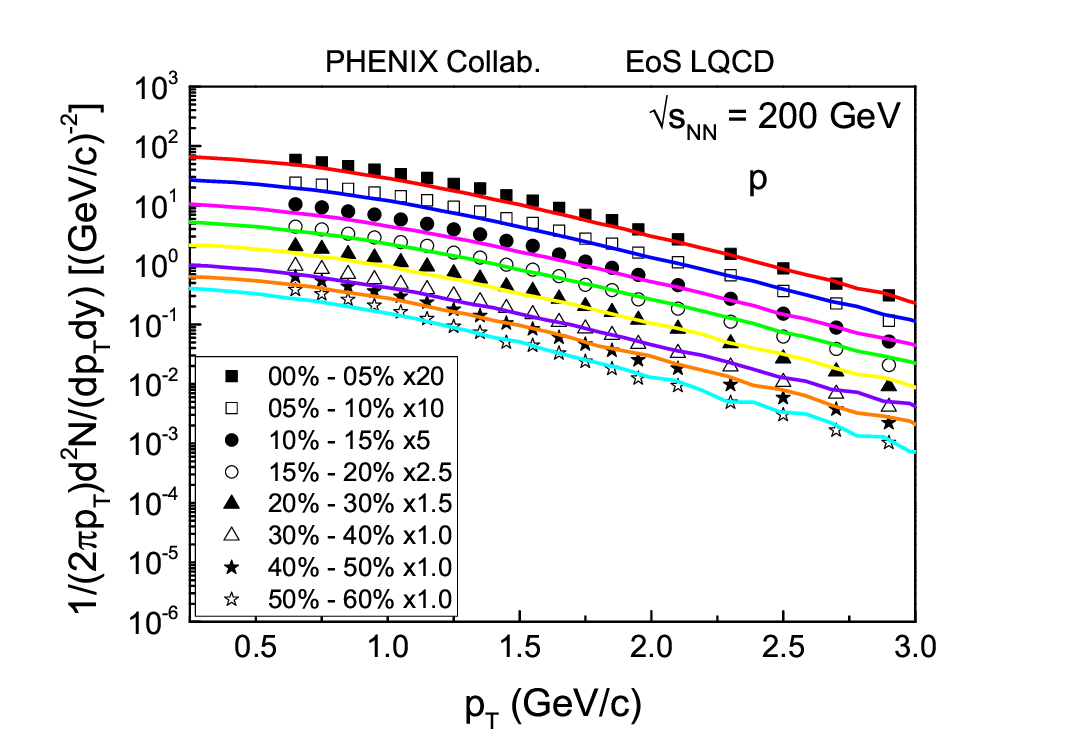}}
\caption{(Color online)
The $p_T$ spectra for identified particles are shown for different centrality windows, corresponding to the FOS, CEP and LQCD EoS's, for Au+Au collisions at 200 GeV.
The data are from the PHENIX Collaboration.}
\label{pt-spectra-pid}
\end{figure*}

There are two free parameters in the present calculations, namely, an overall normalization factor for the IC and the thermal freeze-out temperature.
The former is determined to reproduce the $dN/d\eta$ yields of all charged particles, while the latter is adjusted to the slope of the transverse momentum spectra of all charged particles.
For a given EoS, the freeze-out temperature is therefore determined respectively for each centrality windows of Au-Au collisions at 130 and 200 GeV.
The resulting values of the freeze-out temperatures are presented in Tables \ref{tf130} and \ref{tf200}, and the corresponding transverse momentum spectra of all charged particles are shown in Fig.~\ref{pt-spectra}.
For any centrality window other than those listed in the Table, spline interpolation is used.
Subesquently, observables such as $p_T$ spectra of identified particles, flow parameters $v_2$, $v_3$ and $v_4$ and two pion interferometry are evaluated and compared with the data.
In our calculations, the same IC generated by NeXuS and freeze-out criterion are used in all cases.
An input IC consists of the energy and baryonic density in the co-moving frame and the flow velocity at the initial $\tau = \tau_0 = 1.0$ in hyperbolic coordinates.
We make use of nearly 8000 NeXuS events for each centrality window.
Balancing between a good statistics and efficiency, only 200 events are used for the calculation of particle spectra and two pion interferometry, but all the events are used to evaluate the flow coefficients. 
At the end of each event, a Monte-Carlo generator is invoked 100 times for hadronization. 
The latter is evaluated by the Cooper-Frye formula~\cite{hydro-fz-01,sph-review-1}
\begin{equation}
E\frac{dN}{d^3p}=\int_{\Sigma}f(x,p)p^{\mu}d\sigma_{\mu}=\sum_j\frac{\nu_j n_{j\mu}p^\mu}
          {s_j |n_{j\mu}u_j^\mu|}\;f(u_{j\mu}p^\mu)\,,
\label{cooper-frye}
\end{equation}
where $\Sigma$ is a freze-out hypersurface and $f(x,p)$ is a distribution function.
In the SPH representation, the summation $j$ is taken over all the SPH particles,
which should be taken at the points where they cross the hypersurface $T=T_{f}$, and $n_{j\mu}$ is the normal to this hypersurface.

\begin{figure}[!htb]
\centerline{\includegraphics[width=9cm]{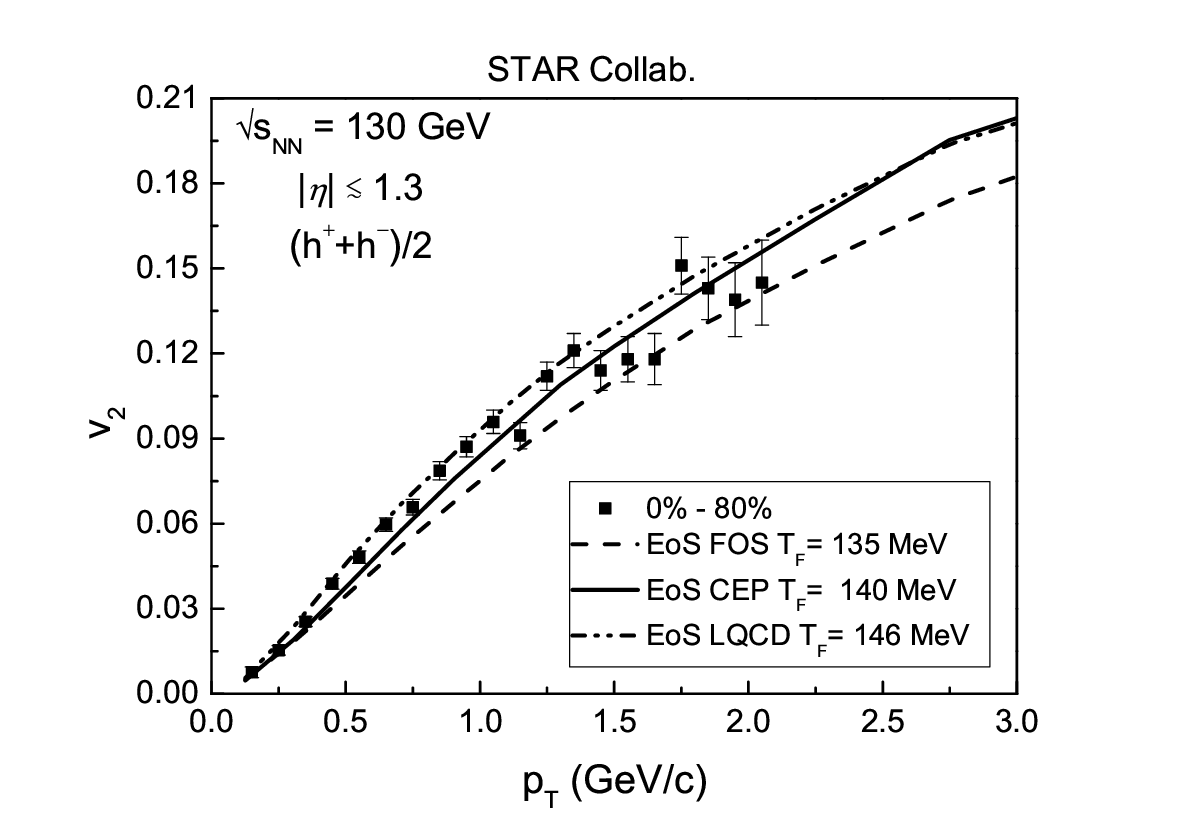} }
\centerline{\includegraphics[width=9cm]{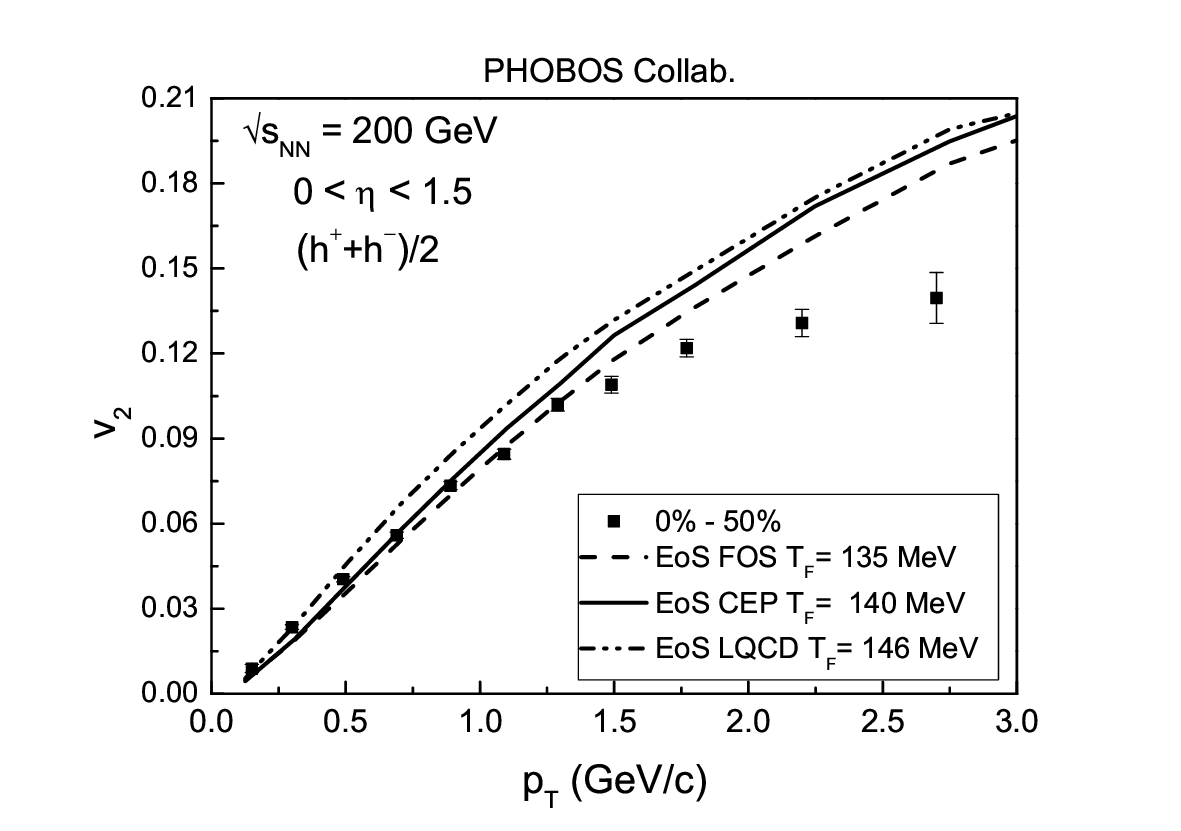} }
\caption{
The elliptic flow coefficient, $v_2$, as a function of $p_T$ is shown for all charged particles, corresponding to the FOS, CEP and LQCD EoS's, for Au+Au collisions at 130 GeV and for 0 - 80\% (top), as well as at 200 GeV and for 0 - 50\% centrality window (bottom).}
\label{v2pt130}
\end{figure}

For illustrating the hydrodynamical evolution, plots for the energy density and entropy density are shown in Figs.~\ref{hydroevolution-eve} and \ref{entropyevolution}.
In Fig.~\ref{hydroevolution-eve} the energy density for a selected random fluctuating event is shown. 
The temporal evolution of the energy density in the transverse plane is calculated considering $\eta = 0$ for the three EoS's, LQCD, CEP and FOS.
The smoothed IC can be obtained in our case by averaging over different fluctuating IC in the same centrality window.
Since it is expected that event by event fluctuations can lead to significant effects on the anisotropic flow components~\cite{sph-v2-1,sph-v2-2} and two particle correlations~\cite{hydro-v3-1,hydro-v3-2}, the calculations in this work are performed using such fluctuating IC.
In the ideal hydrodynamic scenario, the total entropy of the system is conserved.
To see the above results more quantitatively, the snapshots of the entropy density for the same event of Au+Au collision at 200 GeV for 0 - 50\% centrality window are shown in Fig.~\ref{hydroevolution-eve} at typical time instants and the results are depicted in Fig.~\ref{entropyevolution}.
Due to the differences in the EoS's, the same IC may give different total entropy, so the entropy is plotted in percentiles of its initial value, instead of using the absolute values.
\begin{table}[ht]
\centering %
\begin{tabular}{c c c c}
\hline\hline %
  Centrality (\%) & FOS (MeV) & CEP (MeV )& LQCD (MeV) \\ [0.5ex]
\hline
  0 - 5 & 128  & 135  & 145  \\
  5 - 10 & 130 & 136 & 145 \\
  10 - 20 & 132 & 138 & 146  \\
  20 - 30 & 135 & 140 & 147 \\
  30 - 40 & 138 & 142 & 148 \\
  40 - 60 & 143 & 146 & 149 \\ [1ex]
\hline
\end{tabular}
\caption{Centrality (\%) vs. freeze-out temperature $T_{f}$ for Au+Au 130 GeV.}\label{tf130}
\end{table}

\begin{table}[ht]
\centering %
\begin{tabular}{c c c c}
\hline\hline %
  Centrality (\%) & FOS (MeV) & CEP (MeV )& LQCD (MeV) \\ [0.5ex]
\hline
  0 - 6 & 128  & 135  & 145  \\
  6 - 15 & 130 & 137 & 146 \\
  15 - 25 & 134 & 139 & 147  \\
  25 - 35 & 137 & 141 & 147 \\
  35 - 45 & 140 & 143 & 148 \\
  45 - 55 & 143 & 145 & 149 \\ [1ex]
\hline
\end{tabular}
\caption{Centrality (\%) vs. freeze-out temperature $T_{f}$ for Au+Au 200 GeV.}\label{tf200}
\end{table}

\begin{figure*}[!htb]
\centerline{
\includegraphics[width=9cm]{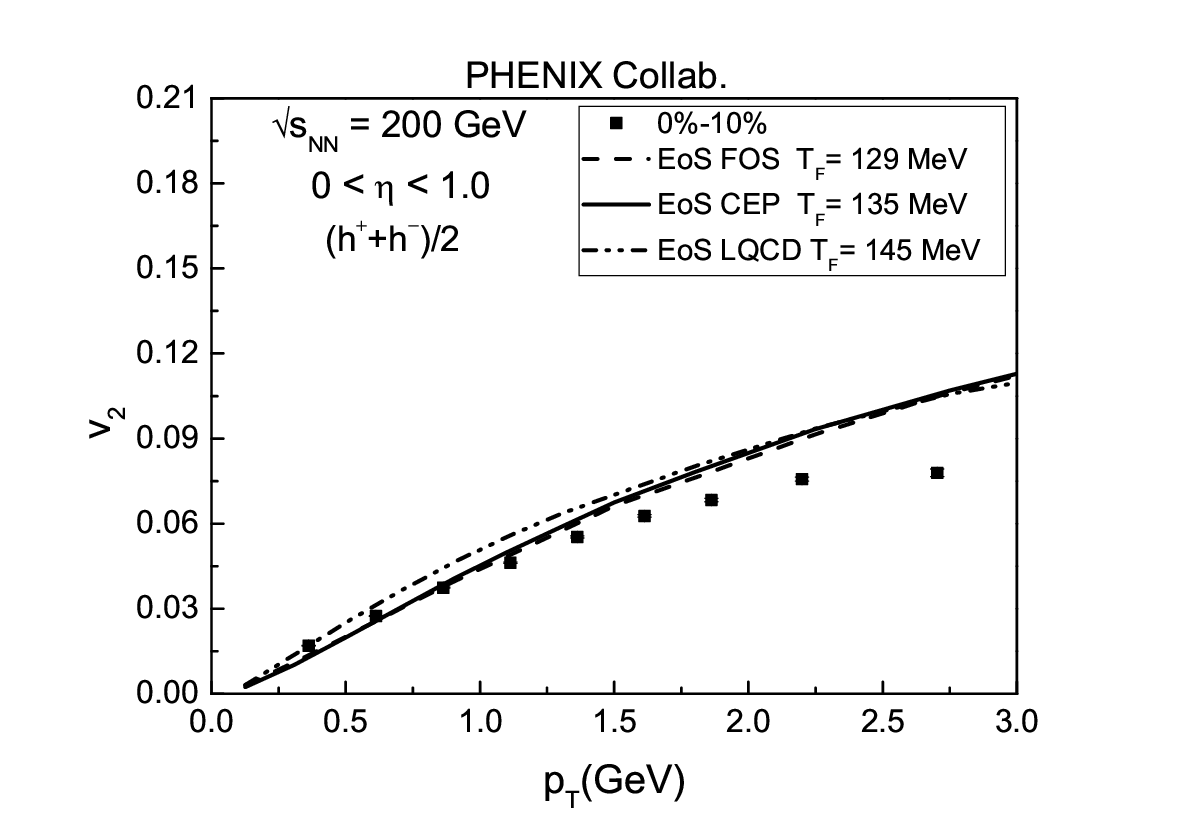}
\includegraphics[width=9cm]{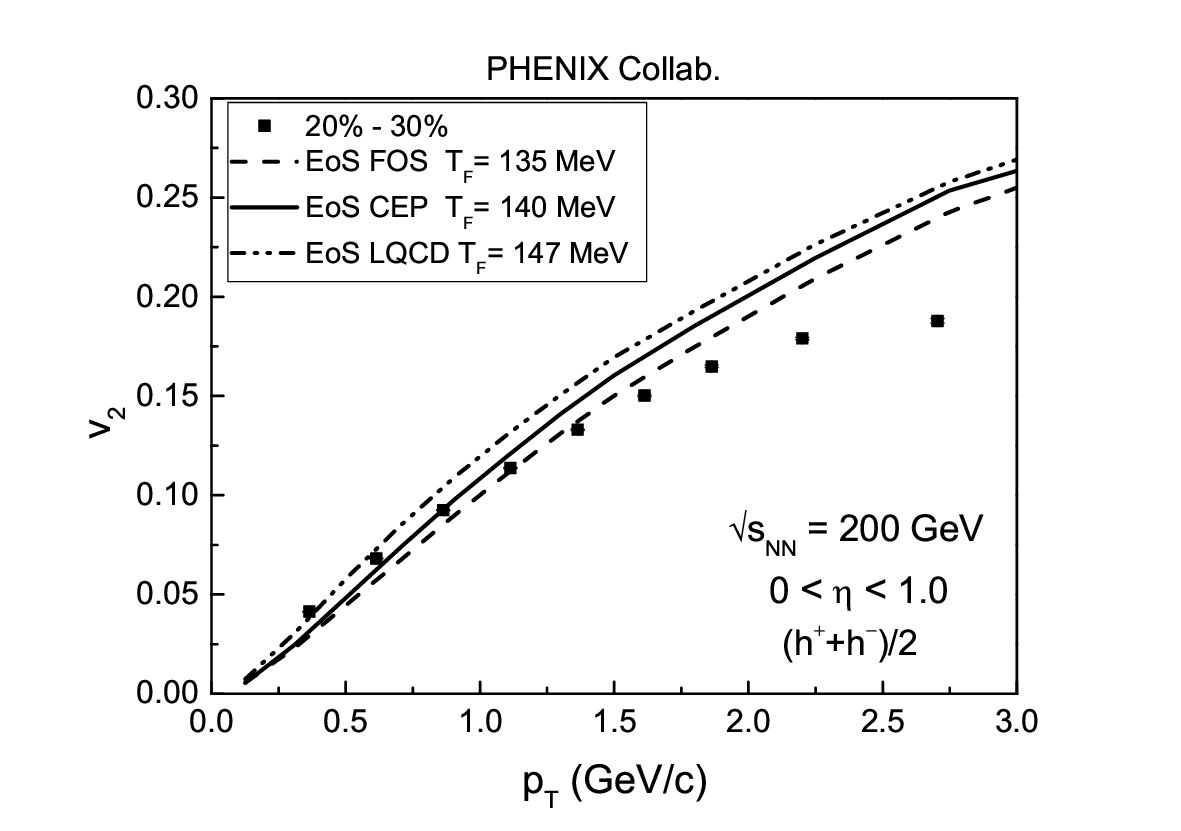}}
\centerline{
\includegraphics[width=9cm]{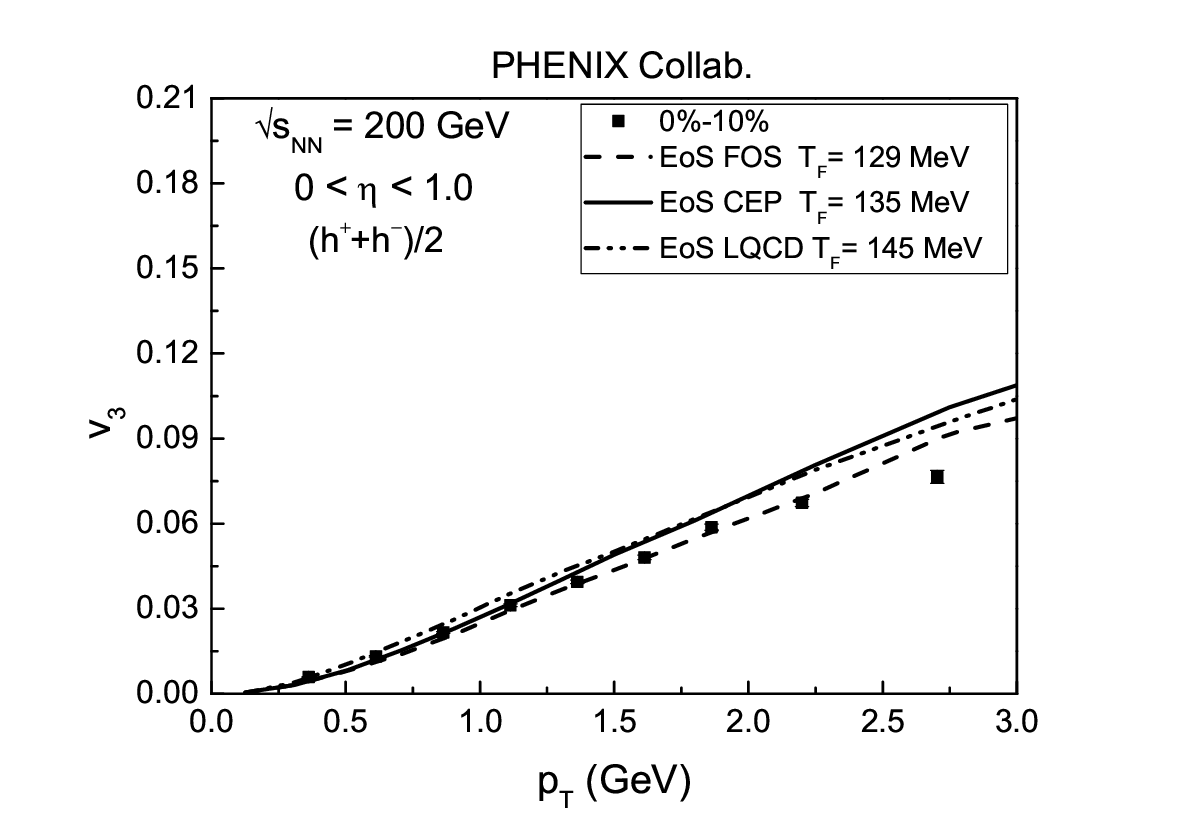}
\includegraphics[width=9cm]{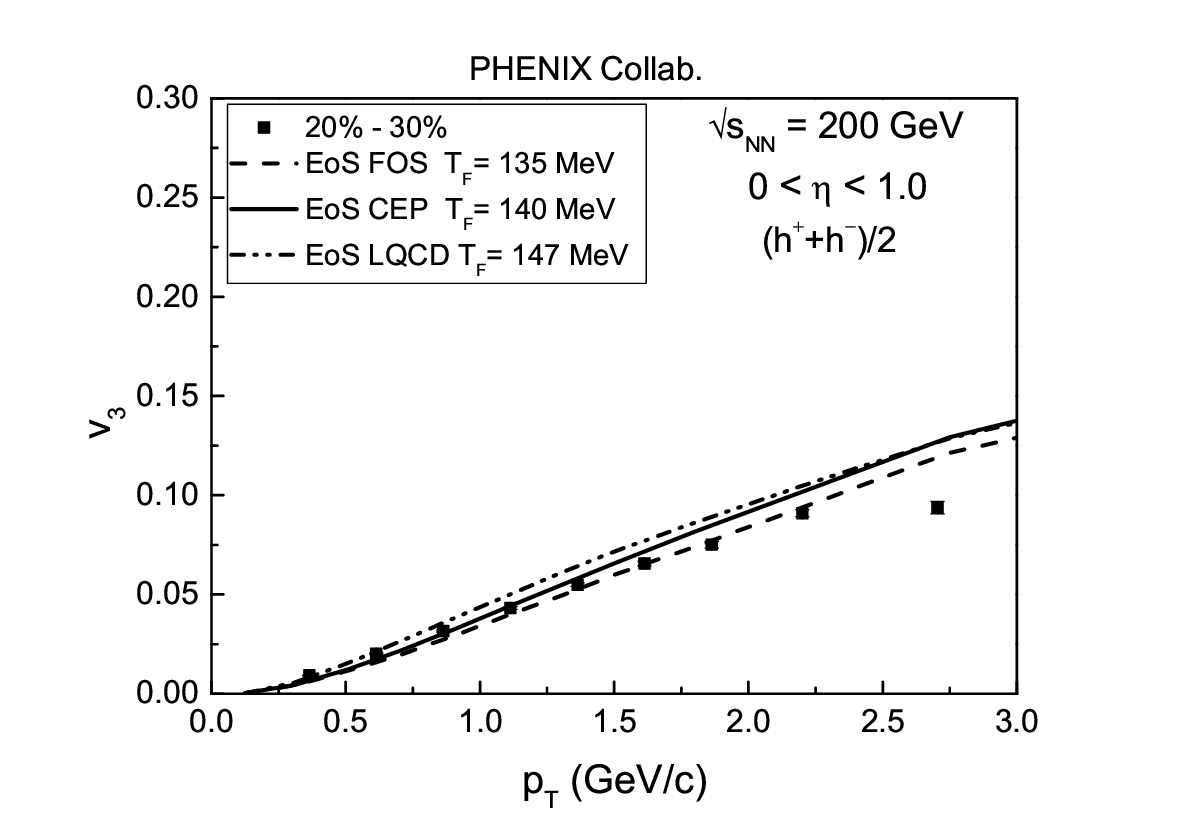}}
\centerline{
\includegraphics[width=9cm]{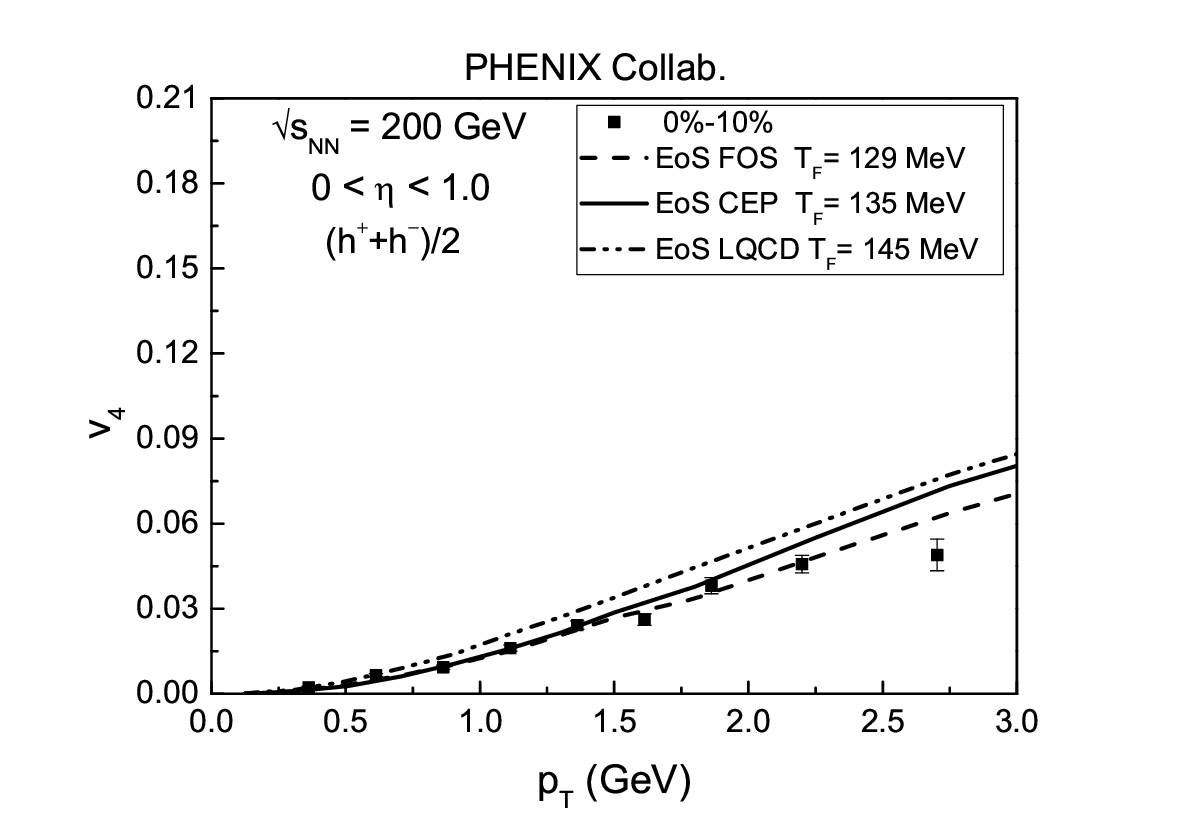}
\includegraphics[width=9cm]{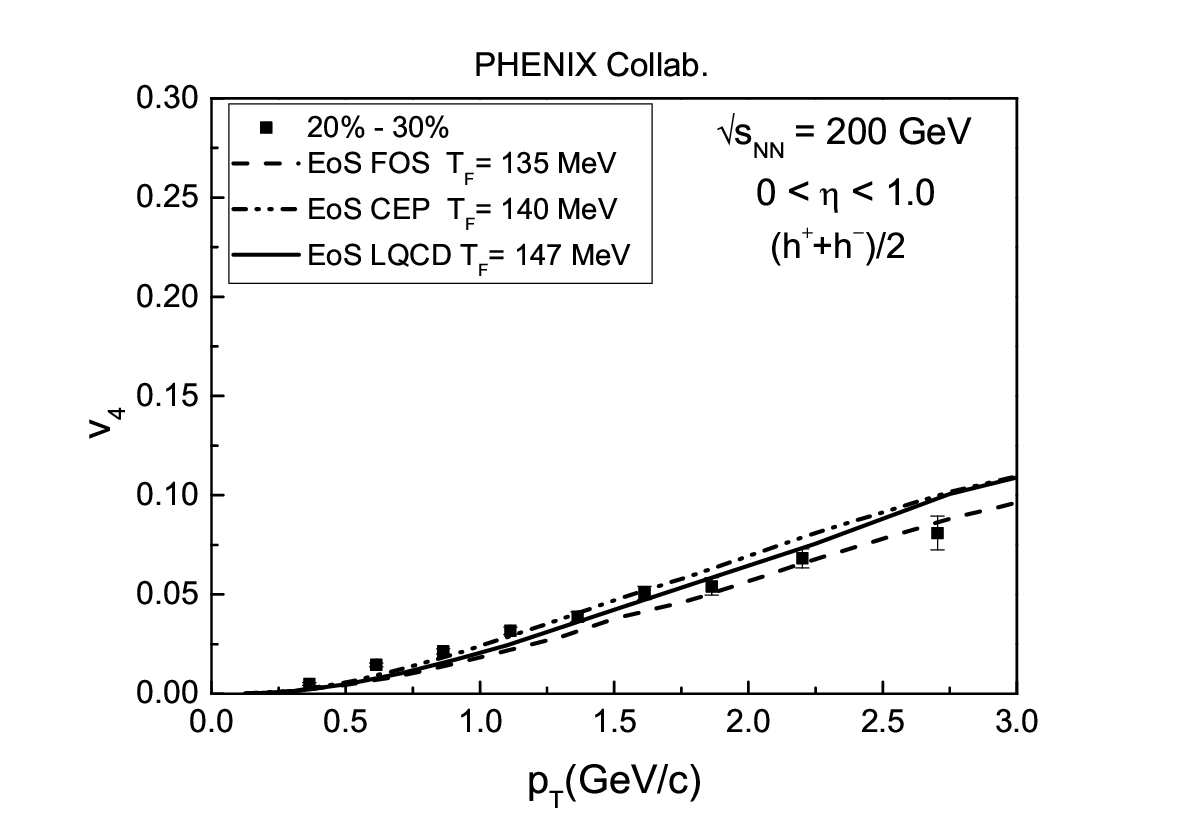}}
\caption{
The anisotropic flow components $v_2$, $v_3$ and $v_4$ are shown as a function of $p_T$ for all charged particles, corresponding to the FOS, CEP and LQCD EoS's, for Au+Au collisions at 200 GeV for the 0 - 10\% (l.h.s.) and for the 20\% - 30\% centrality window (r.h.s.). The data are from the PHENIX Collaboration.}
\label{v2pt200}
\end{figure*}

\begin{figure}[!htb]
\begin{center}
\includegraphics[width=9cm]{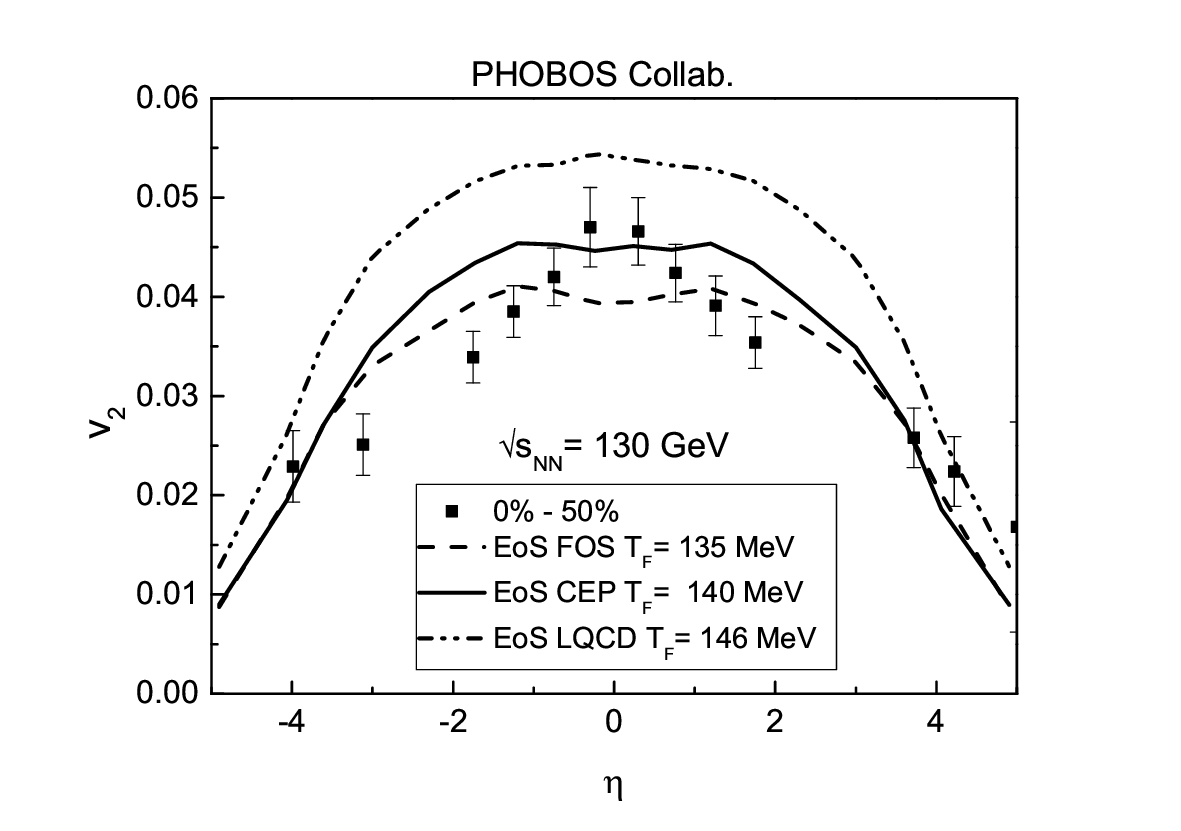}
\includegraphics[width=9cm]{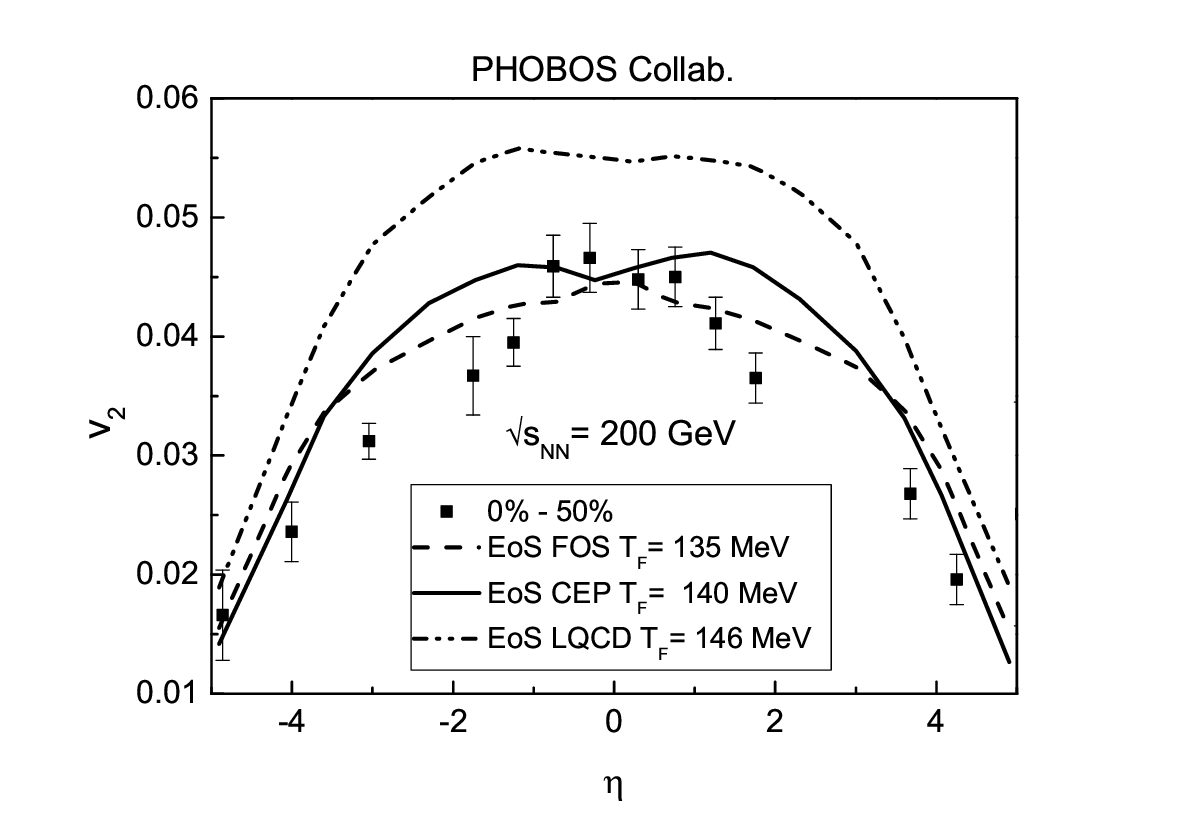}
\end{center}
\caption{
The elliptic flow component $v_2$ is shown as a function of $\eta$ for all charged particles, corresponding to the FOS, CEP and LQCD EoS's, for Au+Au collisions at 130 GeV (top) and 200 GeV (bottom), both for 0 - 50\% centrality window.}
\label{v2eta}
\end{figure}

It can be inferred from the plots that, at both $\sqrt{s} = 130$ MeV and $200$ MeV, the lifetime of the system is the smallest when EoS LQCD is used.
This can be understood using Fig.~\ref{e3pt4}, where for a given energy density, the pressure for LQCD is quite different from those of CEP, FO and FOS.
\begin{figure*}[!htb]
\centerline{
\includegraphics*[width=7.5cm]{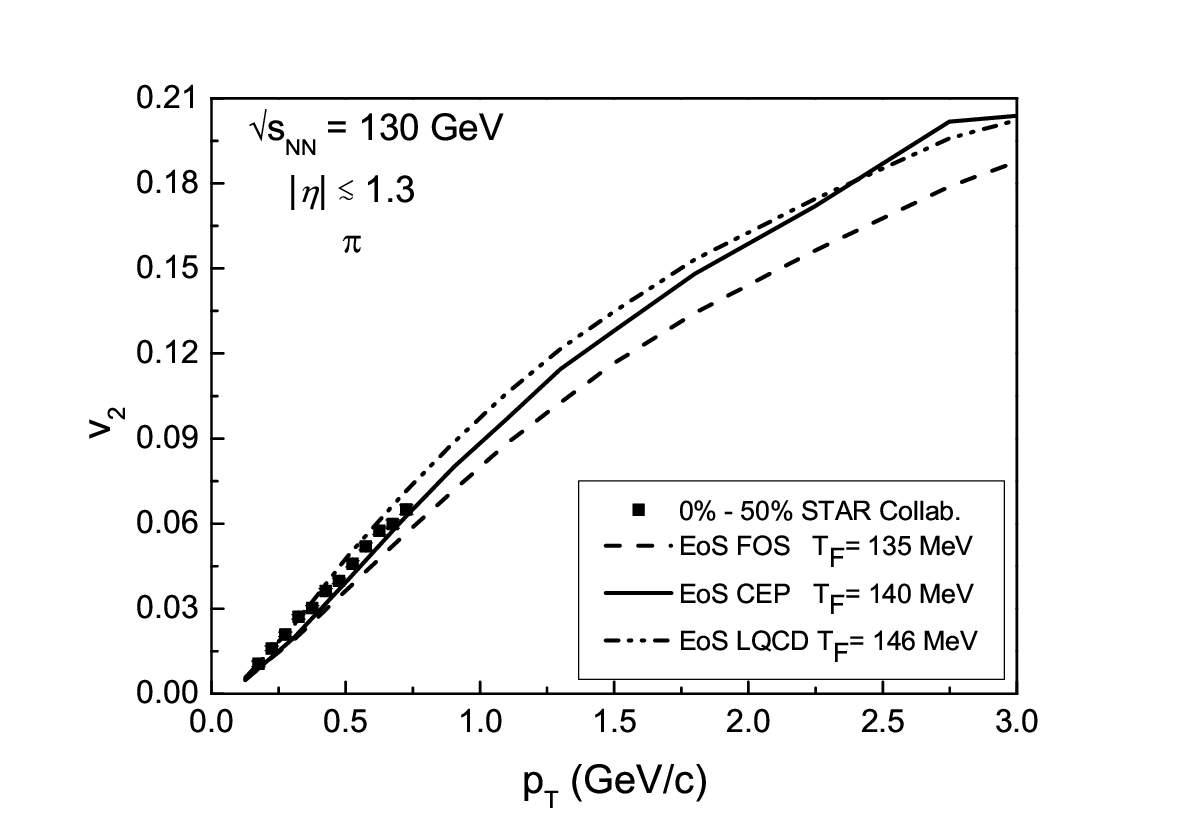}
\includegraphics*[width=7.5cm]{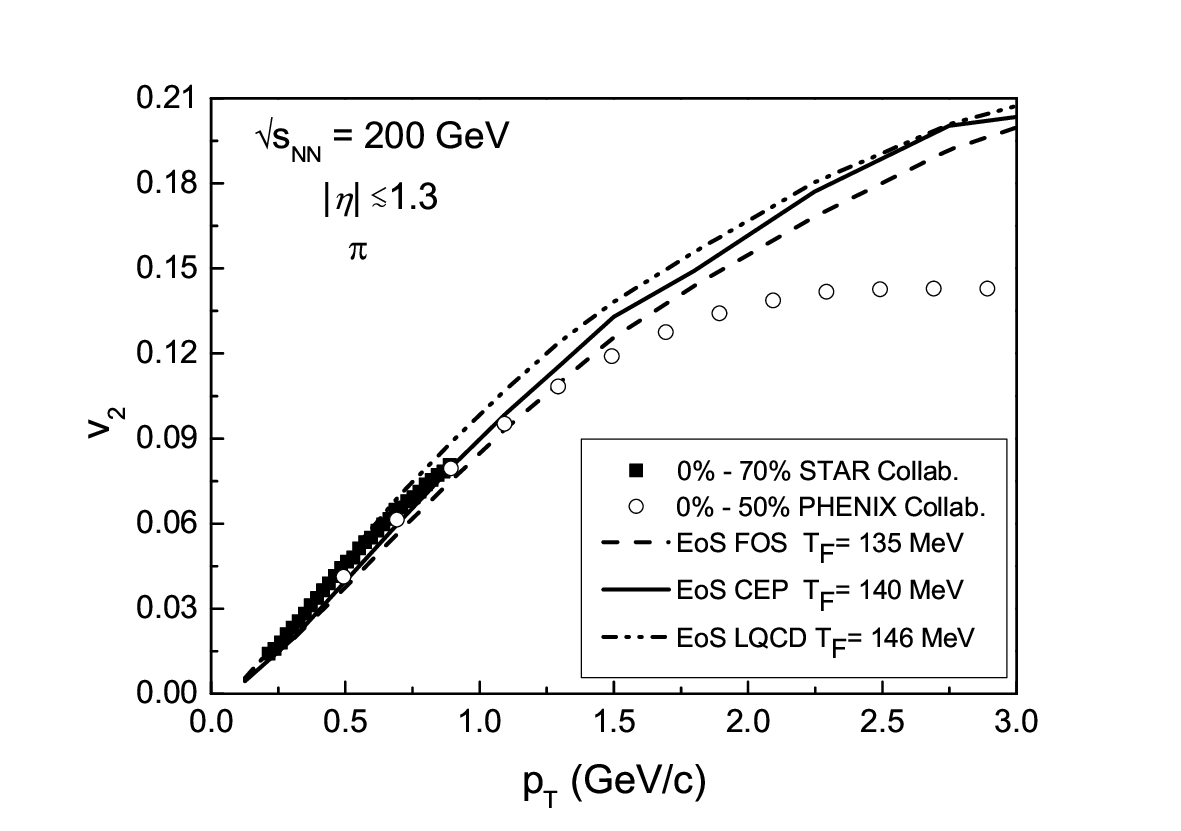} }
\centerline{
\includegraphics*[width=7.5cm]{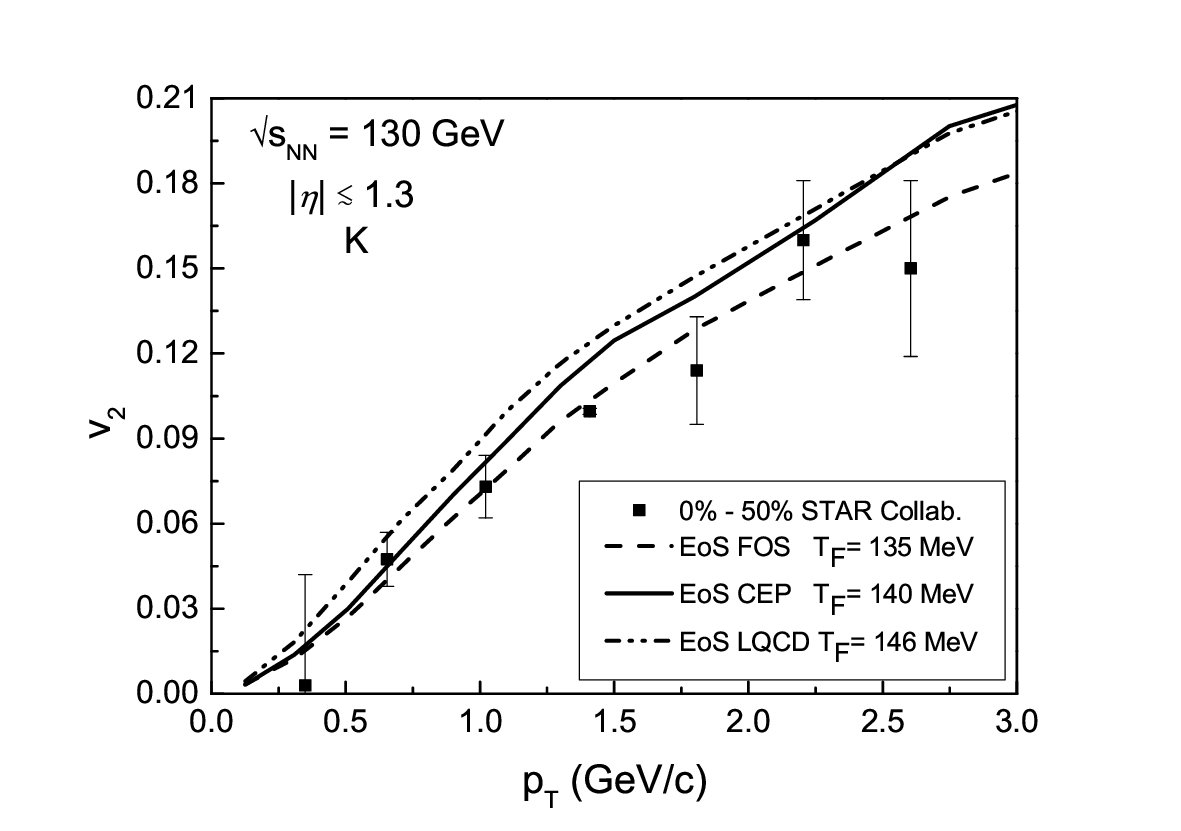}
\includegraphics*[width=7.5cm]{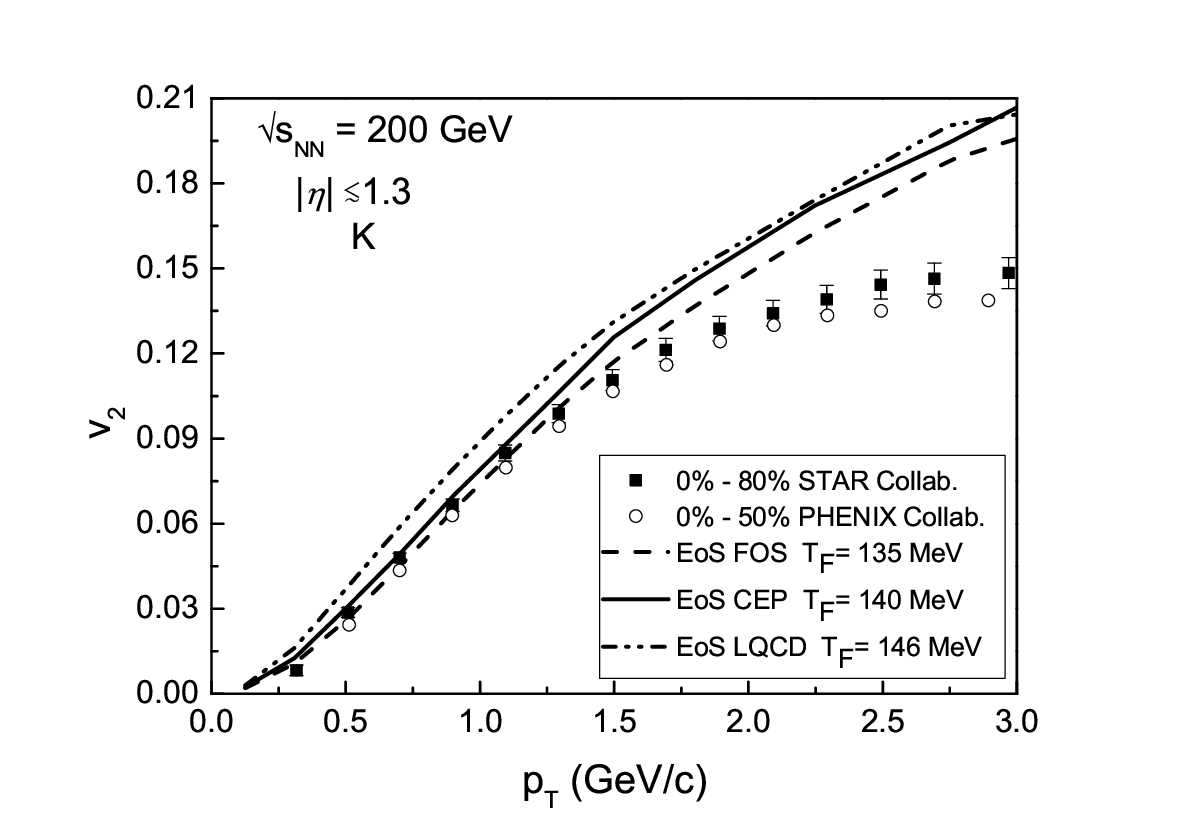} }
\centerline{
\includegraphics*[width=7.5cm]{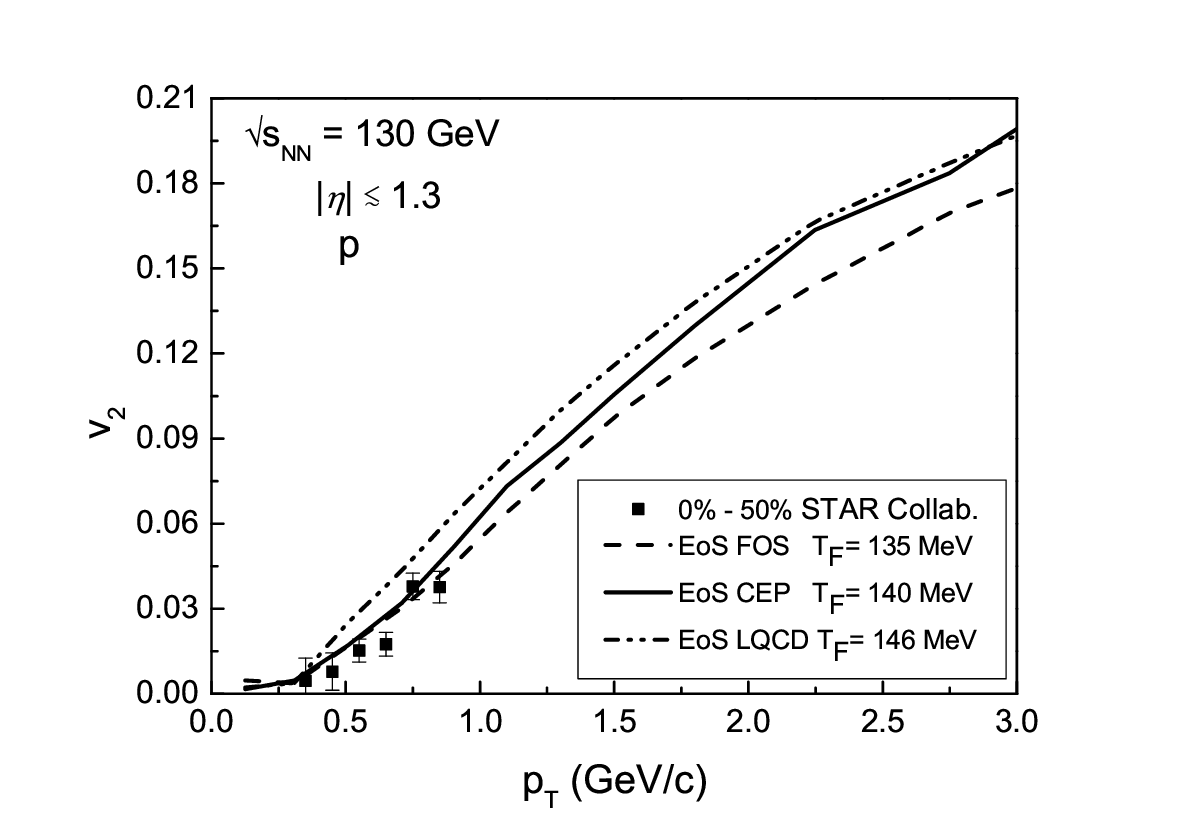}
\includegraphics*[width=7.5cm]{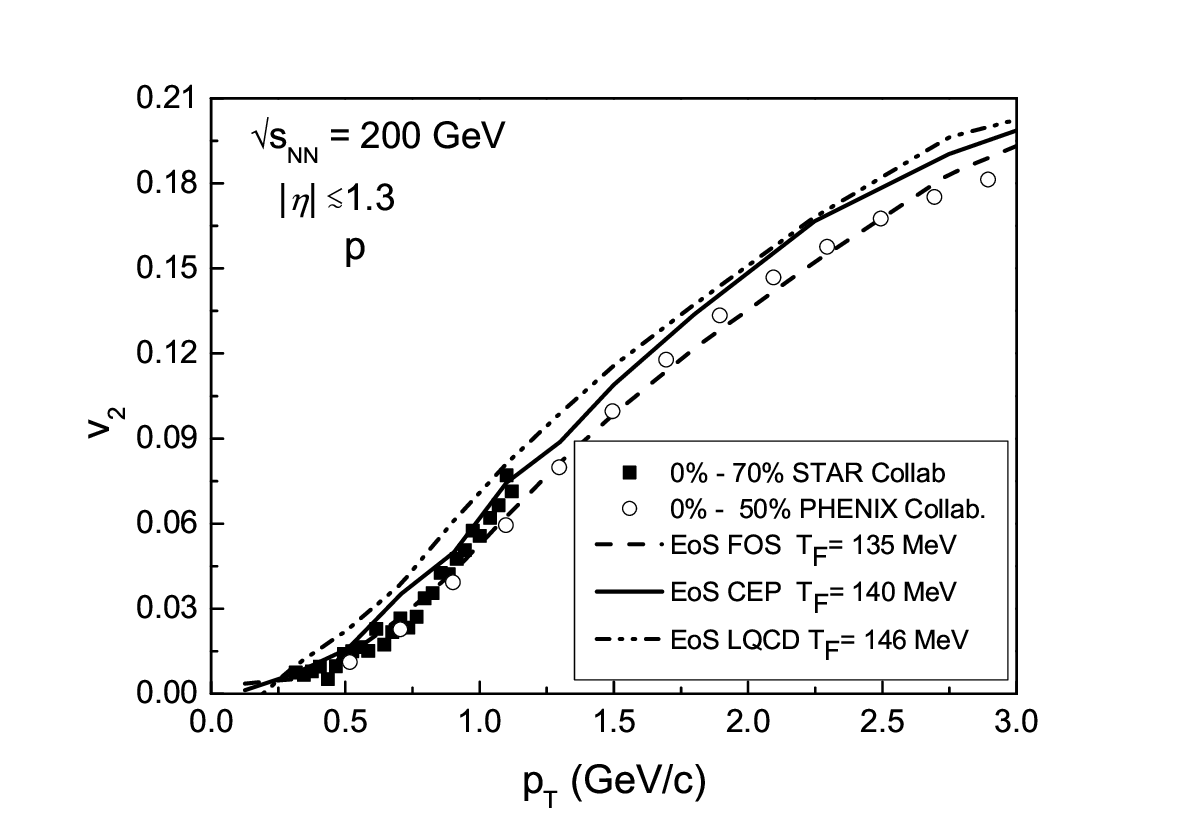} }
\centerline{
\includegraphics*[width=7.5cm]{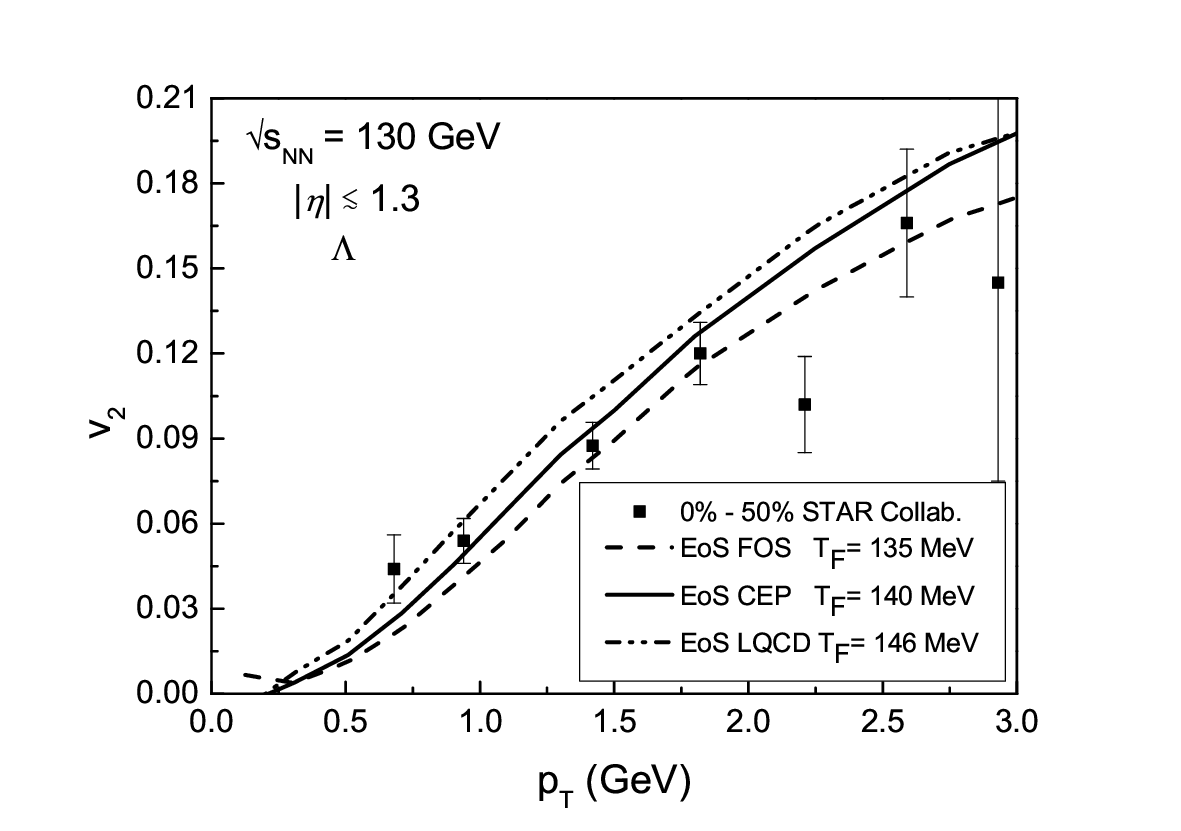}
\includegraphics*[width=7.5cm]{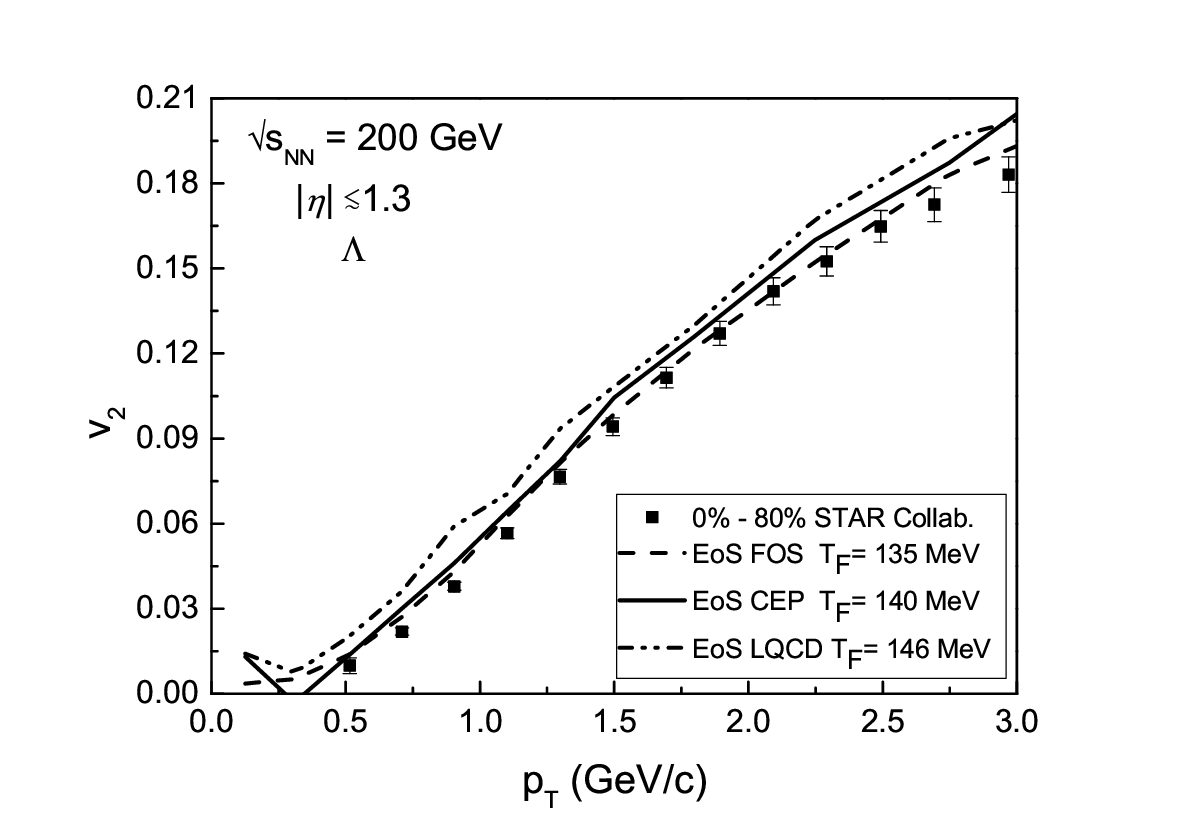} }
\caption{(Color online) $v_2$ vs. $p_T$ for identified particles ($\pi$, K, proton and $\Lambda$) by using different EoS at 130 GeV (left column), and at 200 GeV (right column).}
\label{v2ptid}
\end{figure*}
In particular, for FO with a first order phase transition, the pressure remains unchanged during the transition process meanwhile the system expands.
In the case of CEP and FOS, the phase transition is smoother. 
One observes that for LQCD, the system takes less time to completely freeze-out than for the other three EoS's, which is more evident for the results at 130 GeV for RHIC energy. 
This is probably due to its bigger value of the pressure, as can be seen in Fig.~\ref{e3pt4}, while the differences between CEP, FO and FOS are relatively small.

\begin{figure*}[!htb]
\centerline{
\includegraphics*[width=9cm]{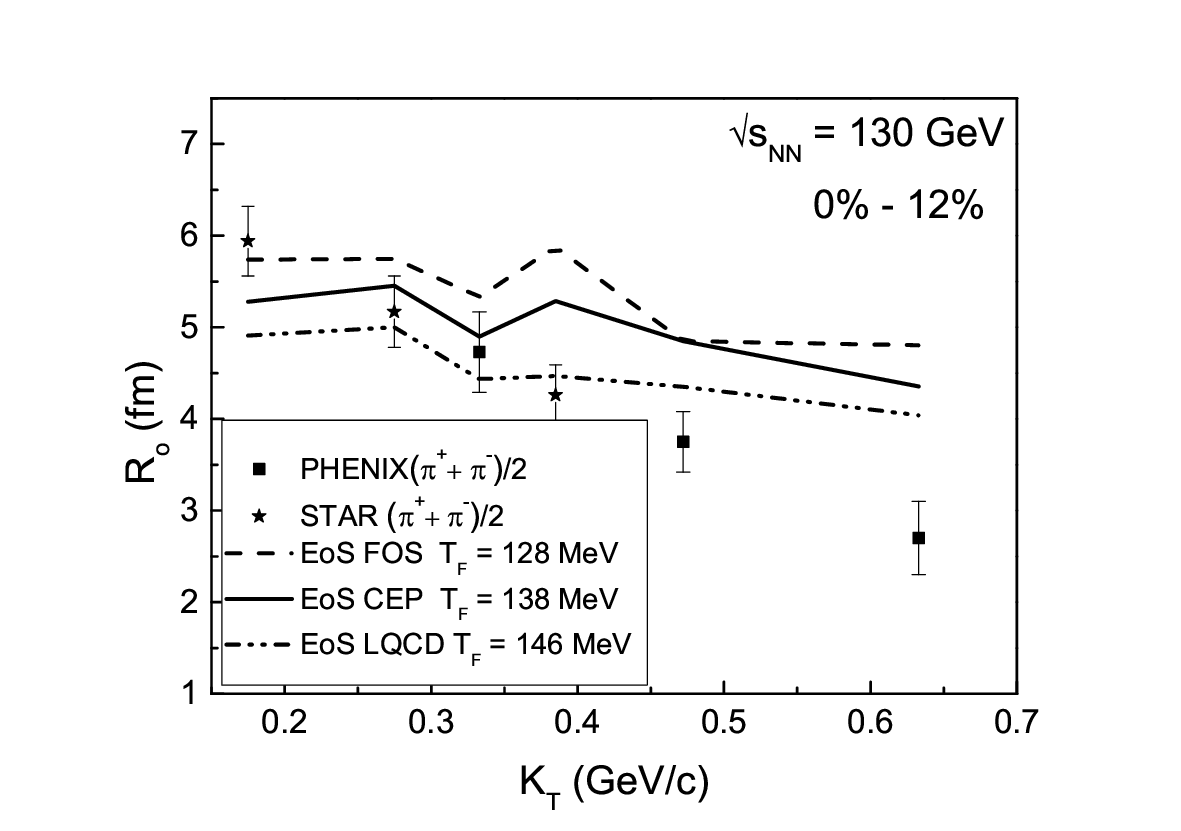}
\includegraphics*[width=9cm]{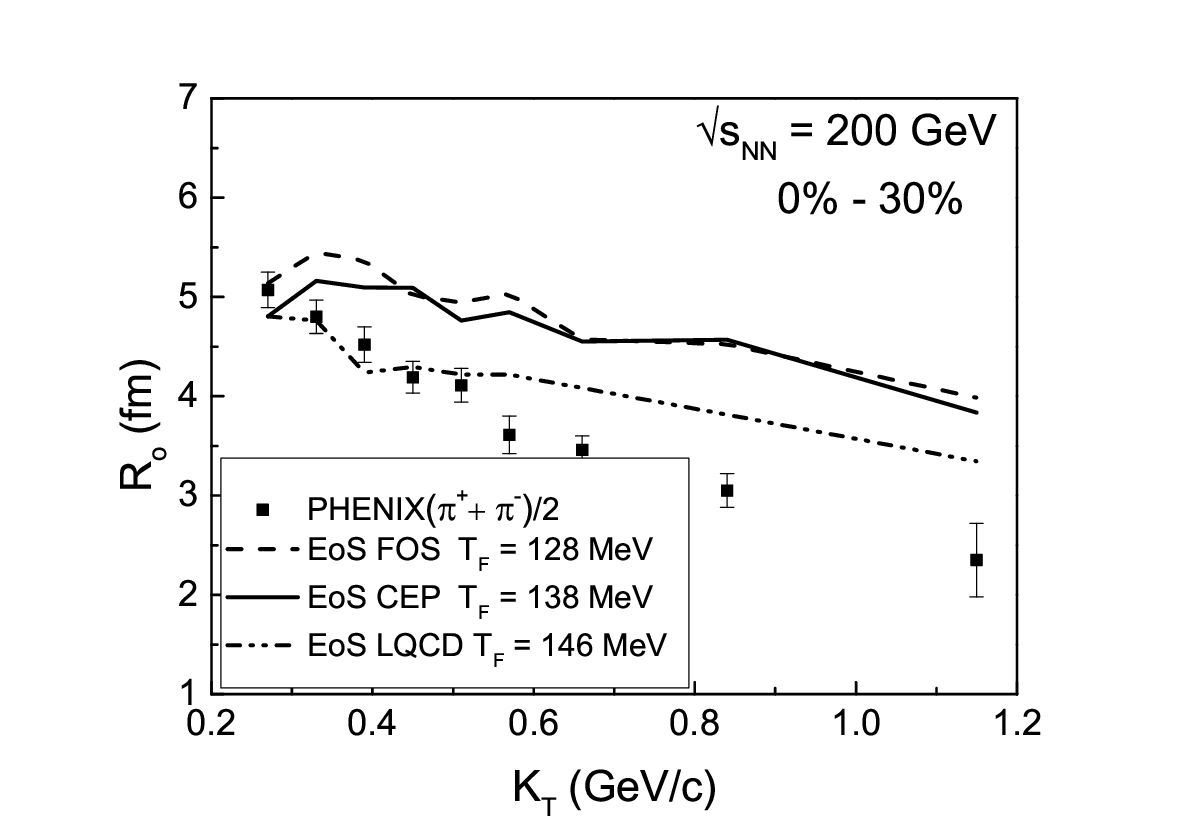} }
\centerline{
\includegraphics*[width=9cm]{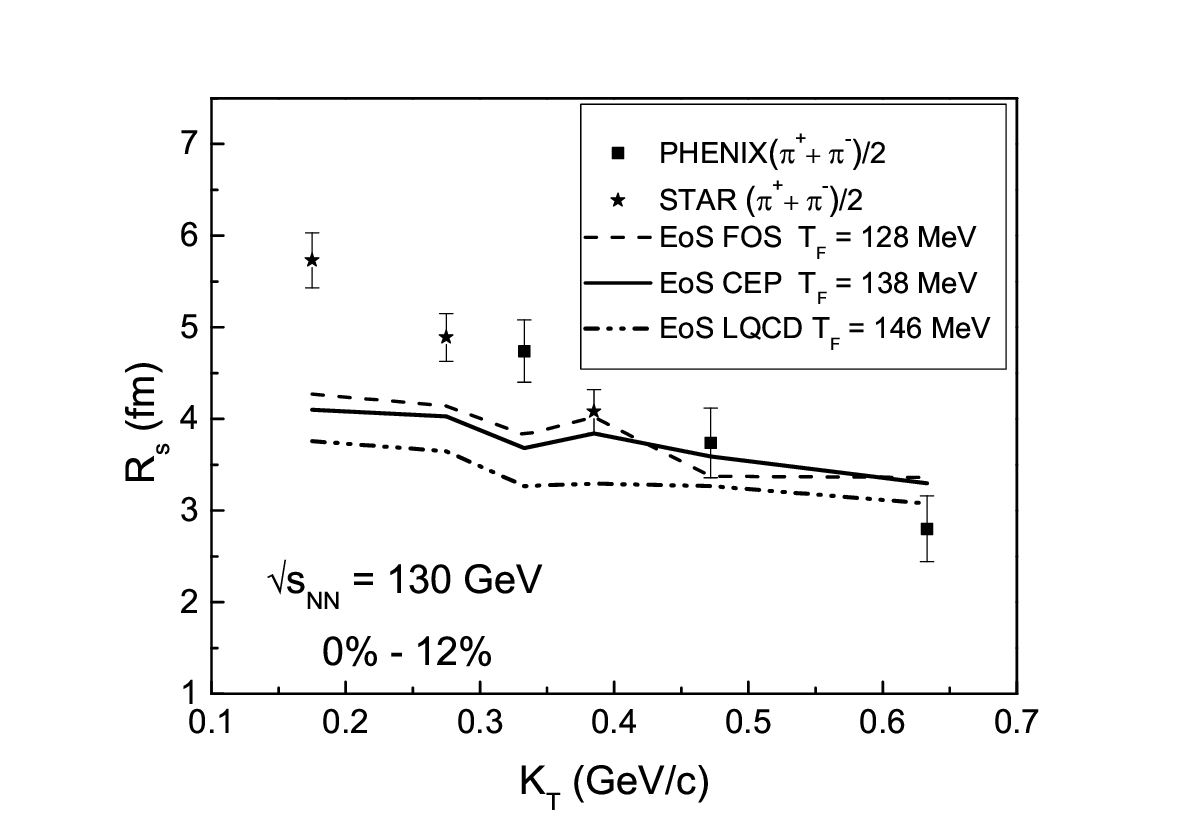}
\includegraphics*[width=9cm]{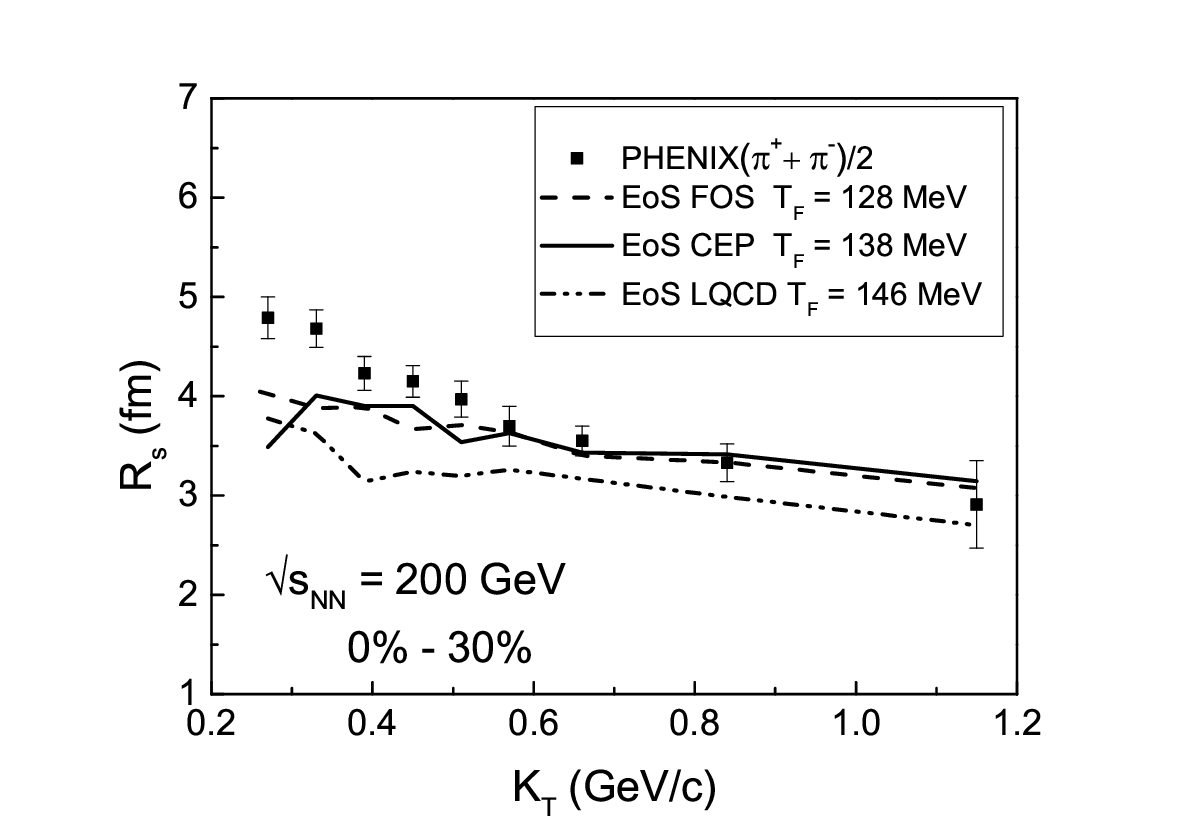} }
\centerline{
\includegraphics*[width=9cm]{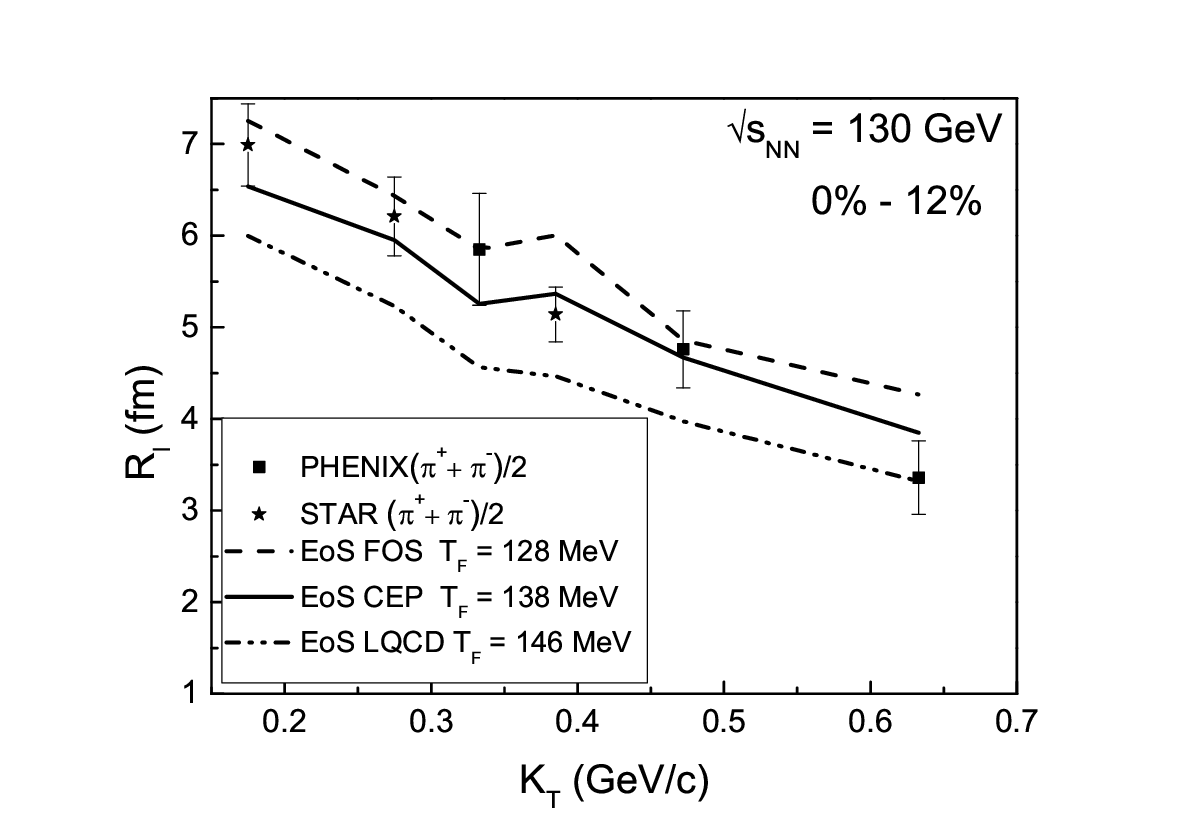}
\includegraphics*[width=9cm]{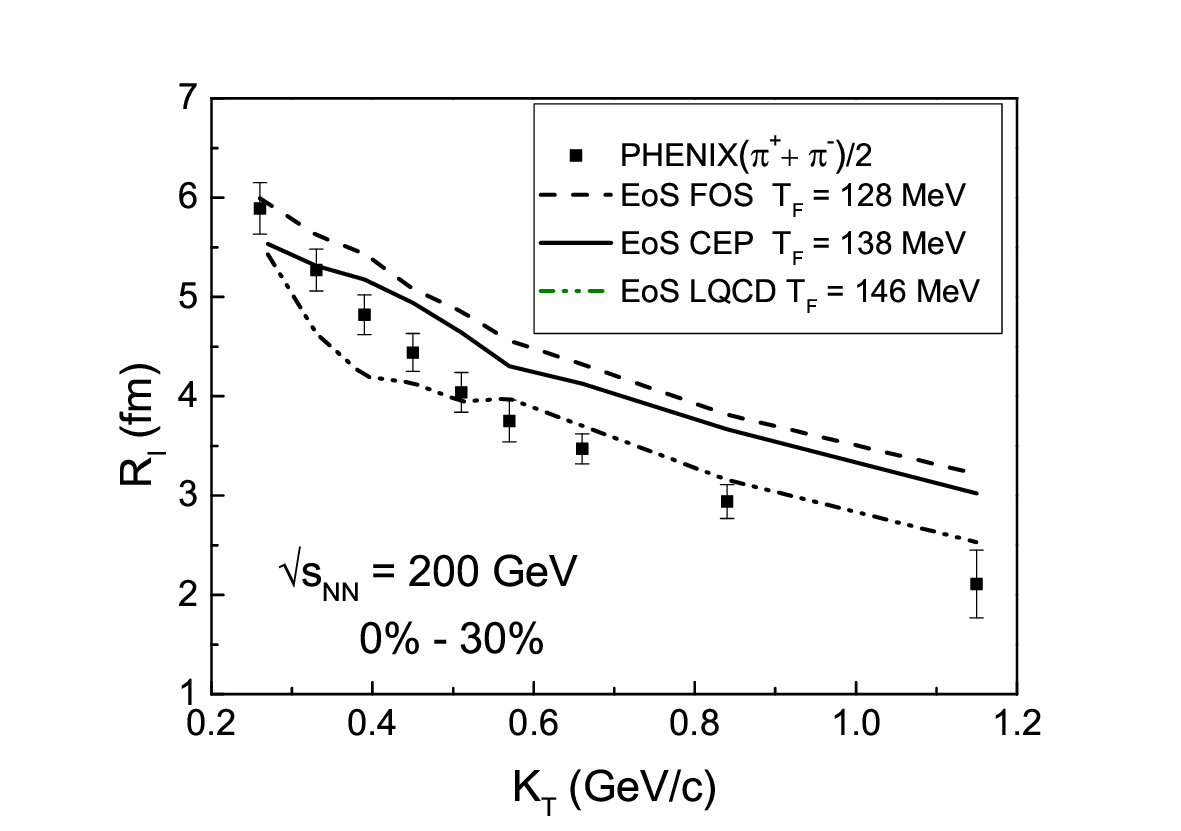} }
\caption{
Calculated HBT radii, $R_o$, $R_s$ and $R_l$. The fit is performed simultaneously for $R_o$, $R_s$ and $R_l$ to a 3D Gaussian function. The results are compared with the PHENIX and the STAR data. The left column: the results for 130 A GeV; the right column: the results for 200 A GeV.}
\label{hbt-cm-lcms}
\end{figure*}

The results of the hydrodynamic simulations for the spectra of identified particles, the flow parameters and the two-pion interferometry are shown in Figs.~\ref{pt-spectra-pid} to \ref{hbt-cm-lcms}.
These results are compared with PHOBOS' Au+Au data at 130~GeV~\cite{RHIC-phobos-spectrum-03} and 200~GeV~\cite{RHIC-phobos-spectrum-02}.

The resulting $p_T$ spectra for identified particles are shown in Fig.~\ref{pt-spectra-pid} where the freeze-out temperatures found in Tables \ref{tf130} and \ref{tf200} are made use of.
For the $p_T$ spectra at 130~GeV, a pseudo-rapidity interval $-1<\eta<1$ is used in the calculations, which is then compared with the STAR data, where the corresponding pseudo-rapidity intervals are $-0.5 < \eta < 0.5$~\cite{RHIC-star-spectrum-01} and $0.5<|\eta|<1$~\cite{RHIC-star-spectrum-02} respectively.
For Au+Au collisions at 200~GeV, the pseudo-rapidity interval $0.2<y<1.4$ used in the calculations of $p_T$ spectra is the same as in the PHOBOS data~\cite{RHIC-phobos-spectrum-01}.
It is noted that the freeze-out temperature is practically insensitive to the incident energy, which is consistent with other studies~\cite{sph-tf-1,RHIC-star-spectrum-03}.
All three EoS's (CEP, FOS, LQCD), reproduce the measured $\eta$ (not shown in the figures) and $p_T$ spectra reasonably well, although some deviations start occurring at $p_T \sim 3$~GeV for peripheral ($\gtrsim$ 40\%) collisions.
We also note that the present model is able to reasonably reproduce the $p_T$ spectra of protons.
This aspect is mostly handled by NeXuS where hadronic decay, as well as final state interaction, are considered after the Cooper-Frye freeze-out.

Next, the results for the harmonic coefficients $v_2$, $v_3$ and $v_4$ are presented.
The collective flow is understood as the response to the initial geometric fluctuations, and can be used to extract information on the eccentricities of the initial conditions and transport properties of the system.
The flow anisotropy coefficients of the angular distribution of the emitted particles are defined as
\begin{equation}
\frac{dN}{d\phi}=\frac{N}{2 \pi} [1+2\sum_{n=1}^{\infty} v_n \cos n(\phi-\Psi_n)],
\label{flow}
\end{equation}
where $N$ is the number of particle created, $\phi$ is the azimutal angle,  $\Psi_n$ the events plane (EP) angles~\cite{hydro-vn-0,RHIC-phenix-v2-3}, defined as 
\begin{equation}
\Psi_n=\frac{1}{n}\mathrm{arctan2}(\langle \mathrm{sin}(n\phi)\rangle, \langle \mathrm{cos}(n\phi)\rangle) .
\label{flow2}
\end{equation}
For $n=2$, $\Psi_2$ is considered the best approximation to the reaction plane, defined by the collision direction and the collision impact parameter, which is not directly accessible experimentally.

Here all the calculations are done by using the event plane method~\cite{RHIC-phenix-v2-3}, and the results for $v_n$ are obtained as a function of pseudorapidity, the transverse momentum. 
For Au+Au collisions, the calculated $v_2$ is shown in Fig.~\ref{v2pt130} as a function of $p_T$ and compared with STAR~\cite{RHIC-star-v2-3} results (top) and PHOBOS~\cite{RHIC-phobos-v2-2} results at 200~GeV (bottom).
Similar calculations of $v_2$ as a function of $\eta$ for Au+Au collisions are carried out at these two energies and compared with data from these two collaborations in Fig.\ref{v2eta}.  
In the calculations of $v_n$ as function of $p_T$ at 200~GeV, only particles in the interval $-1.0<\eta<1.0$ are considered, whereas the intervals $-3.9<\eta<-1.0$ and $1.0<\eta<3.9$ are used to evaluate the event planes in the forward and backward directions.
The results are compared to the data from the PHENIX collaboration~\cite{RHIC-phenix-vn-01,RHIC-phenix-vn-02} in Fig.~\ref{v2pt200}.
When calculating $v_n$ as function of $p_T$, a cut in pseudo-rapidity $|\eta|< 1.3$ was implemented. 
There is no momentum cut in the calculations of $v_2$ as a function of $\eta$.
In both cases, the freeze-out temperatures are from the Tables \ref{tf130} and \ref{tf200}.
In our calculations, the obtained $v_n$ is corrected by using the event plane resolution Res{$\Psi_n$}, in accordance with the two sub-event method~\cite{RHIC-phenix-v2-3}.
We use 1000 events for 0 - 10\% centrality window, 8000 events for 20\% -30\% and the Monte Carlo generator was invoked 500 times for each given IC.

The calculations are also carried out for identified particles as well.
In the left column of Fig.\ref{v2ptid}, the results are shown for $v_2$ vs. $p_T$ for pions, kaons and protons from Au+Au collisions at 130~GeV. 
The calculations were performed in the 0 - 50\% centrality window.
The experimental data are from the STAR collaboration~\cite{RHIC-star-v2-2,RHIC-star-v2-5}.
In the right column, we present the corresponding results of $v_2$ vs. $p_T$ for these same particles at 200~GeV. 
The calculations were done for the 0 - 50\% centrality window and are compared to the STAR data corresponding to 0 - 80\%~\cite{RHIC-star-v2-6} and 0 - 70\%~\cite{RHIC-star-v2-1}, as well as to the more recent data from PHENIX~\cite{RHIC-phenix-vn-03} for 0 - 50\% centrality window.

From these plots, it is seen that at small $p_T$ the measured elliptic flow coefficients can be reasonably well reproduced by using all three EoS's.
Overall, different EoS's give roughly similar results, and all of them fail to describe the kaon data at 200 GeV when $p_T$ increases beyond $\sim $~2~GeV, probably reflecting the limitations of the ideal hydrodynamics employed in this study.
It can be seen that the results of LQCD are slightly different from those of CEP and FOS for $p_T < 1$ GeV.
This can be understood as follows. For an EoS featuring a first order phase transition, such as FO, the pressure gradient vanishes when the system enters this region.
In the present case, although CEP and FOS describe smooth phase transitions, their properties are more similar to that of FO than to those of LQCD. For LQCD, it could be inferred from Fig.~\ref{e3pt4} that the pressure gradient is bigger than those from other EoS's in the high-temperature region ($T \geq 0.3$ GeV). 
As a result, in the case of LQCD, the initial spatial eccentricity of the system would be transformed into momentum anisotropy with the biggest amplified magnitude at high temperature as well as during the hadronization process.
At 200 GeV, higher initial temperatures could be achieved, and consequently, the matter in evolution spends more time in the QGP phase where most eccentricity is developed. Therefore the asymptotic behavior of LQCD at high temperature plays an increasingly important role, which makes it distinct from all other EoS's.
On the other hand, at a lower incident energy, 130~GeV, the system develops relatively more anisotropy in the transition region. Therefore the properties of the phase transition become more important. 
This makes the curve of CEP to become closer to that of LQCD.

The last part of this paper is dedicated to the investigation of the influence of different EoS's on two-particle quantum statistical correlations, also known as GGLP~\cite{hbt-03,hbt-01,hbt-sinyukov-01,hydro-hbt-01} effect, which is the analog in the high energy collisions realm of the Hanbury-Hanbury Brown and Twiss (HBT) effect~\cite{hbt-02}. 
The HBT/GGLP effect is used to estimate the apparent size of the systems formed in relativistic heavy ion collisions at freeze-out, i.e., the lengths of homogeneity. 
These systems expand rapidly while exhibiting strong collective effects. 
Due to their collectivity, particles with similar momenta are likely emitted from the same spatial region. 
In consequence, the two-particle correlation function depends not only on the pair relative momentum but also on the average pair momentum.
Therefore, the measured particle pairs with smaller average momentum would contribute to larger lengths of homogeneity.
In this context, the radius parameter measured by HBT effect is sensitive to both the size of the source region and to the average momentum of the emitted particle pair.
In the calculations, we consider the emitted particle pairs ($k_1, k_2$) are created in the center of mass reference system (CM) and transformed to ($k_1^*, k_2^*$) the local co-moving reference system (LCMS).
The latter satisfies $K_L^*=0$ with $\vec{K}^*=1/2(\vec{k}_1^*+\vec{k}_2^*)$.
The correlation funtion of two identical particle is calculated as
 \begin{equation}
C_2 (\vec{p}_1,\vec{p}_2)= 1+ 
\lambda \exp{\left\{-(R_l^2 q_l^2 + R_o^2 q_o^2 + R_s^2 q_s^2)/(\hbar c)^2\right\}}, \label{lcmsfunc}
\end{equation}
where $p_1$ and $p_2$ are the momenta of the particle pair, $\lambda$ is the chaotic parameter, $R_s$, $R_o$ and $R_l$ are the HBT radii and $q_s$, $q_o$ and $q_l$ are the corresponding momentum components~\cite{hydro-hbt-01}.
The results for the lengths of homogeneity of pions are shown in Fig.\ref{hbt-cm-lcms} and compared with the data from STAR~\cite{RHIC-star-hbt-1} and PHENIX~\cite{RHIC-phenix-hbt-1,RHIC-phenix-hbt-2}\footnote{We note that recent data from PHNIX Collaboration~\cite{RHIC-phenix-hbt-3} is not available on its website, and therefore in this work, we adopt the previous results by PHENIX and STAR.}.
In general, the obtained HBT radii are in reasonable agreement with the data, but they overestimate the data for $R_o$ at large $K_T$ while underestimate those for $R_s$ at small $K_T$.

Among the three different EoS's, the LQCD gives mostly smaller values in comparison with FOS and CEP.
This is consistent with what is shown in Fig.~\ref{entropyevolution}.
Since the lifetime of the system expansion for LQCD EoS is smaller, and this results in smaller HBT radii.
On the basis of the above results, one concludes that the observed differences due to different EoS's are small in size.

\section{Conclusions and perspectives}
\label{sec: concl}

A systematic study of the role of the EoS on hydrodynamical evolution of the system is carried out and discussed here.
By adopting the parameters tailored to each specific EoS, which consist of an overall renormalization factor and the freeze-out temperatures, the particle spectra, anisotropic flow coefficients and two-pion HBT correlations are calculated by NEXSPheRIO code. 
The calculations cover a wide range of centrality windows at two different RHIC energies. 
All EoS's are found to successfully reproduce the particle spectra and elliptic flow at small $p_T$ region.
The hydrodynamical evolution of the system is affected by the EoS, which consequently leads to some small differences observed in collective flow.
In the case of HBT radii, all the EoS's give reasonable but not exact description of the data in all $p_T$ range.

The effect of EoS on was also carried out recently by other authors~\cite{eos-hydro-01,eos-hydro-02,eos-hydro-03}.
By using the experimental data as a constraint, it is shown that the resulting possible EoS's are consistent with those from LQCD~\cite{eos-hydro-02}.
LQCD was extended to consider finite chemical potential~\cite{eos-pasi-04} where the pressure is expanded in terms of chemical potentials by a Taylor expansion coefficients which are parameterized and compared to those obtained from lattice simulations.
The existing results indicate that LQCD reproduces results closer to the data than the other EoS's investigated here.
Other factors, such as different types of IC, fluctuations in the IC, viscosity, etc., should also be considered carefully. 
In this case, the Bayesian statistics might be utilized~\cite{hydro-analysis-01}.
The EoS's with finite baryon/strangeness density also provide results with small but observable differences from other EoS's that assume zero chemical potentials.
Additionally, the time evolution, as well as momentum anisotropy, are shown to be affected. 
Also, various Lattice QCD groups have updated the EoS results, especially for those at finite baryon density, in the past few years~\cite{lattice-07,lattice-08,lattice-09,lattice-10,lattice-11}.
Therefore, it is interesting to introduce an EoS which considers finite chemical potential while reproduces the lattice data at high temperature and zero baryon density region. 
Such an EoS may be employed to consistently study physical systems over a broad range of densities and temperatures.

In our present approach, the resulting difference between different EoS's is not significant.
From a hydrodynamical viewpoint, to obtain a more distinct result, especially concerning the existence of a critical point, one may construct specific EoS's which focus on the properties of the assumed critical point.
In other words, EoS's otherwise identical except in the vicinity of the critical point, specifically, there might be a smooth crossover or the end point to a first order phase transition.
The EoS in question might be elaborated by using a quasiparticle model~\cite{sph-eos-4}.
By appropriately adjusting the collision energy for a suitable value of baryon density, the trajectory of the temporal evolution determined by adiabatic curves in the phase diagram might pass right through the location of the critical point.
Subsequently, by evaluating observables, especially those proposed for the beam energy scan program~\cite{RHIC-star-bes-01,RHIC-star-bes-02,RHIC-star-bes-03,RHIC-star-bes-04}, such as particle ration, multiplicity, as well as $p_T$ fluctuations, harmonic flow coefficients, and dihadron correlations, it is more likely to obtain a more significant difference. 
We plan to carry out such studies in the near future.

Hydrodynamics assumes local thermal equilibrium, based on which the dynamical properties of the hot and dense system are expressed in terms of the EoS.
The calculations carried out in this work show that the results are weakly dependent on the EoS.
One of the possible reasons for this is that all the observables studied here involves event-average procedure, and thus may not necessarily represent genuine (real event-by-event) non-linear hydrodynamic evolution. 
That is, if the event averaged correlations keep a strong linear relation to the corresponding ones in the initial density fluctuations~\cite{hydro-vn-3}, the non-linear dynamics can only manifest in the event-by-event distribution of the correlations mentioned above and not in the event averaged value. 
It is interesting to note that the transport models such as AMPT or PHSD have shown to have a similar properties as viscous hydrodynamic calculations~\cite{ampt-4,ampt-5,hydro-review-06}, but when we look into on real event-by-event basis, a state close to the local thermal equilibrium appears only in a tiny space-time domain during the dynamical evolution (Ref.~\cite{hydro-review-06,phsd-01}).  
To clarify up to what extent the genuine event-by-event hydrodynamics is valid, it is required a new set of observables which are sensitive to the non-linear evolution of the system. 
Works in this direction are under consideration.

\section{Acknowledgments}
\label{sec:Ack}

The authors are thankful for valuable discussions with Wojciech Florkowski and Tam\'as Cs\"org\"o.
We gratefully acknowledge the financial support from Funda\c{c}\~ao de Amparo \`a Pesquisa do
Estado de S\~ao Paulo (FAPESP), Funda\c{c}\~ao de Amparo \`a Pesquisa do Estado de Minas
Gerais (FAPEMIG), Conselho Nacional de Desenvolvimento Cient\'{\i}fico e Tecnol\'ogico (CNPq),
and Coordena\c{c}\~ao de Aperfei\c{c}oamento de Pessoal de N\'ivel Superior (CAPES).
A part of this work has been done under the project INCT-FNA Proc. No. 464898/2014-5.

\bibliographystyle{h-physrev}
\bibliography{references_qian}{}

\end{document}